\documentclass[sigconf,nonacm]{acmart}

\usepackage{listings}
\usepackage{makecell}
\usepackage{multirow}
\usepackage{tabularx}
\usepackage{array}
\usepackage{adjustbox}
\usepackage{placeins}
\usepackage{xspace}
\usepackage{etoolbox}

\newif\ifshowpending \showpendingfalse

\def\tsc#1{\csdef{#1}{\textsc{\lowercase{#1}}\xspace}}
\tsc{WGM}
\tsc{QE}

\begin{document}

\title{LLM-Generated Counterfactual Stress Scenarios for Portfolio Risk Simulation via Hybrid Prompt-RAG Pipeline}

\author{Masoud Soleimani}
\email{m.soleimani@ieee.org}
\affiliation{%
  \institution{Department of Information Engineering, University of Pisa}
  \city{Pisa}
  \country{Italy}}

\renewcommand{\shortauthors}{Soleimani}

\begin{abstract}
We develop a transparent, auditable large language model (LLM) pipeline for
macro--financial stress testing. Structured prompting, with optional retrieval of
country fundamentals and news, emits G7 macroeconomic scenarios (GDP growth,
inflation, policy rate), which are converted into deviations from an IMF World
Economic Outlook baseline and translated into portfolio losses through a
three-channel PCA factor model. The macro-to-factor mapping is \emph{stipulated}
rather than estimated, so every tail-risk multiple is conditional on stated
assumptions. Across 2{,}420 accepted scenarios from two model families and three
runs, plus an unanchored-severity run and eight risk-layer ablations, we find
three things. First, severity is inherited from scenario design, not generated by
the model: prompt variant dominates every variance decomposition, removing the
prompted severity bands makes GPT-5.6-luna's recession $1.6$~pp milder,
apparent country heterogeneity largely reflects the institutional baseline, and
retrieval and news matter only where the prompt leaves magnitude free. Second,
numerical severity and narrative severity are only weakly coupled. Third,
GPT-5.6-luna follows its own arithmetic but anchors on prompted bands, whereas
Llama-3.1-8B is four to seven times more dispersed and contradicts its own stated
changes in $90\%$ of cases. A repeat run differs at the scenario level yet agrees
on aggregate multiples to within $0.007$ times. LLM scenario generators are best
used for narrative and sector elicitation around a prescribed severity path, not
as autonomous calibrators of stress magnitude.
\end{abstract}


\maketitle

\section{Introduction}

Macroeconomic stress testing is central to financial stability analysis and bank supervision \citep{taskinsoy2022stress}. Stress tests articulate adverse but plausible macroeconomic conditions to assess vulnerabilities, determine capital adequacy, and inform policy design. Regulatory authorities such as the Federal Reserve and the ECB provide top--down narratives, while financial institutions implement internal frameworks based on historical replays, econometric models, or Monte Carlo simulations \citep{thiemann2021measuring, Guarda2012MVAR}. Yet, despite their institutional importance, traditional stress--testing pipelines face persistent challenges. First, they struggle to represent low-frequency disruptions such as pandemics, supply-chain failures, energy shocks, or geopolitical fragmentation \citep{Borio2014StressTesting, Carney2015Horizon} that fall outside econometric training windows. Second, scenario design remains manually intensive \citep{Baer2023ClimateScenarios, Aikman2024ECBStress} and difficult to scale across jurisdictions or portfolios. Third, econometric systems are often slow to adapt to real--time information \citep{Pesaran2005RealTime}, limiting responsiveness in fast-moving macro--financial environments.

\medskip
\noindent\textbf{Large Language Models (LLMs)} offer a promising complement. Their ability to synthesise structured narratives from heterogeneous sources has been demonstrated in policy analysis \citep{safaei2024end} and macroeconomic forecasting \citep{Carriero2024MacroLLM}, and their use in finance is surveyed below. For stress testing, LLMs can rapidly generate country-specific macroeconomic scenarios while remaining interpretable to human analysts. However, unconstrained generation poses well-known risks: hallucination, numerical drift, internal inconsistency, and limited reproducibility \citep{ji2023towards, huntsman2024prospects, staudinger2024reproducibility, Tan2025ConsistentLLMs, Tonmoy2024HallucinationSurvey}. These issues motivate hybrid architectures with explicit grounding, structure, and diagnostics.

\medskip
\noindent In this paper we develop and evaluate a transparent, retrieval--optional pipeline for macro--financial scenario generation. Our system couples structured country profiles with optional news retrieval, prompting GPT-5.6-luna and Llama-3.1-8B-Instruct to emit machine-readable macro shocks (GDP, inflation, interest rates). These shocks are then translated into portfolio losses through a linear factor channel, enabling direct computation of scenario-induced VaR and CVaR multiples relative to historical and econometric baselines. The design preserves narrative flexibility while enforcing numerically stable, auditable outputs.

\paragraph{Motivation.}
Traditional econometric stress tests struggle to scale or update quickly, while fully generative LLM approaches lack governance guarantees. Our aim is to bridge these approaches: retaining the interpretability and auditability of structured stress-testing frameworks while exploiting the adaptability and expressiveness of modern LLMs.

\medskip
\noindent\textbf{The contributions of this paper are:}
\begin{enumerate}
    \item A fully auditable Prompt--RAG pipeline for macro--financial scenario generation, with structured JSON outputs and optional grounding via country profiles and news retrieval.
    \item A comprehensive G7 experiment comprising \emph{840 intended scenarios per model}
    (7 countries $\times$ 30 prompt variants $\times$ 4 retrieval configurations),
    of which 838/838/744 survive the schema and coherence gate for the primary
    GPT-5.6-luna run, the repeat GPT-5.6-luna run, and Llama-3.1-8B-Instruct,
    respectively.
    \item A consistent macro$\rightarrow$portfolio mapping using a linear factor channel, enabling computation of VaR/CVaR multiples relative to historical bootstrap, EWMA, and GARCH(1,1)--t baselines.
    \item Extensive diagnostics including scenario plausibility checks, dispersion analysis, cross-run stability, cross-country consistency metrics, and ANOVA variance decomposition, showing that prompt design dominates risk variance, retrieval has no detectable effect, and news has a small effect only where the prompt leaves magnitude free.
    \item A pre-declared ablation grid---four generation conditions and eight risk-layer variants---covering severity anchoring, shock source, macro--equity calibration (including an attempted empirical estimate), stress-covariance estimator, factor-extraction universe and the policy-rate baseline (Table~\ref{tab:ablations}).
    \item A complete reproducibility and governance layer: primary run modes, hash-verified artifact manifests, and explicit snapshotting to ensure replayability across models and retrieval configurations.
\end{enumerate}

From a practitioner perspective, the pipeline is best viewed as a
``scenario generator'' for risk committees: given a fixed regulatory or
internal baseline, the LLM produces a menu of country-specific,
narrative-rich shocks that can be screened, edited, and selectively
added to an institution's scenario library, rather than replacing
existing frameworks.

\section{Related Work}
\label{sec:related}

\paragraph{Large language models in finance.}
Early NLP applications used domain lexicons and linear models, but struggled with contextual nuance.~\cite{mishev2020evaluation, Tetlock2007InvestorSentiment, LoughranMcDonald2011Liability}.  
Transformer architectures and later, large language models (LLMs), closed that gap \cite{Araci2019FinBERT, Yang2020FinBERT}. The survey of Xing et al.\ documented over 40\% accuracy gains relative to bag-of-words baselines on stock-return prediction tasks~\cite{xing2018natural}.  
Domain-specific pre-training further boosts performance: BloombergGPT (50B parameters) outperformed general models by up to 15 pp on 14 financial NLP benchmarks~\cite{wu2023bloomberggpt}, while the open-source FinGPT project emphasises continual web-scale fine-tuning for reproducibility~\cite{liu2023fingpt}.  
LLMs also show strong zero-shot capabilities: ChatGPT improves short-horizon equity-return forecasts from headlines~\cite{LopezLira2024PredictiveEdge}, and GPT-4 can generate analyst-style reports that meaningfully influence professional investors~\cite{takayanagi2025can, Gruver2023ZeroShotTS, Carriero2024MacroLLM}.

\paragraph{Machine-learning stress testing.}
Traditional macro–financial stress tests rely on vector autoregressions and hand-crafted macroeconomic shocks \cite{Guarda2012MVAR, Aikman2024ECBStress, AlfaroDrehmann2009MacroStress}.  
Machine-learning variants broaden the factor set and automate aspects of scenario design \cite{FlaigJunike2022ScenarioGen, Petropoulos2022DeepStress}.  
Packham shows that PCA and autoencoders uncover latent factors that better explain tail losses than regulator-supplied shocks~\cite{packham2024risk}.  
Bueff et al.\ propose counterfactual generators for credit portfolios that yield interpretable “closest-possible’’ stress scenarios~\cite{bueff2022machine}.  
Moffo demonstrates that gradient-boosted trees outperform linear satellite models in the CCAR environment, while noting their limited narrative interpretability~\cite{moffo2024machine}.  
These approaches operate on structured numerical data and do not generate full narrative macroeconomic scenarios, despite the growing emphasis on explainable stress testing \cite{Nallakaruppan2024XAI, Sun2025ExplainableEnsemble}.

\paragraph{Retrieval-augmented generation (RAG)}
LLMs hallucinate when extrapolating beyond their training distribution.  
Retrieval-augmented generation (RAG) mitigates this by conditioning model outputs on retrieved ground-truth documents~\cite{lewis2020retrieval, Asai2024SelfRAG, Yan2024CorrectiveRAG}.  
WebGPT augmented GPT-3 with a browser and citation mechanism, raising factual accuracy to 74\% on long-form QA~\cite{nakano2021webgpt}.  
Atlas matched GPT-3-175B performance using an 11B model paired with a neural retriever and frozen language model~\cite{izacard2023atlas}.  
In finance, Zhang et al.\ integrate a news-article retriever with GPT to improve sentiment accuracy by up to 48\% over non-retrieval baselines~\cite{zhang2023enhancing}.  
We are not aware of prior work combining retrieval-augmented generation with macroeconomic stress-scenario generation and downstream portfolio risk translation, though we note that this is a claim about our search rather than an exhaustive one: our review covered the finance and NLP venues listed above plus forward and backward citation chasing from the works cited here, and did not follow a pre-registered systematic protocol.

\paragraph{Position of this study.}
This work bridges these strands.  
We combine IMF fundamentals with optional news retrieval using a MiniLM--FAISS
retriever to ground prompts for GPT-5.6-luna and Llama-3.1-8B-Instruct.  
The hybrid Prompt--RAG pipeline produces structured JSON macro shocks that
(i) are statistically plausible relative to historical and econometric
VaR/CVaR baselines, (ii) exhibit cross-country heterogeneity, and
(iii) remain interpretable for supervisory-style review.  

We have not found an end-to-end pipeline that couples retrieval-augmented LLM
macro scenarios with explicit portfolio-level VaR/CVaR translation, subject to
the search caveat above. We do not benchmark directly against
CCAR/EBA or bank internal scenario libraries; instead, we position our
framework as a complement to those tools, providing a scalable way to
generate additional, grounded narratives around existing macro baselines.
Our factorial G7 evaluation quantifies how models, retrieval choices, and
macro narratives influence tail-risk outcomes.

\section{Methodology}
\label{sec:methodology}

\begin{figure*}[t]
  \centering
  \includegraphics[width=0.95\linewidth]{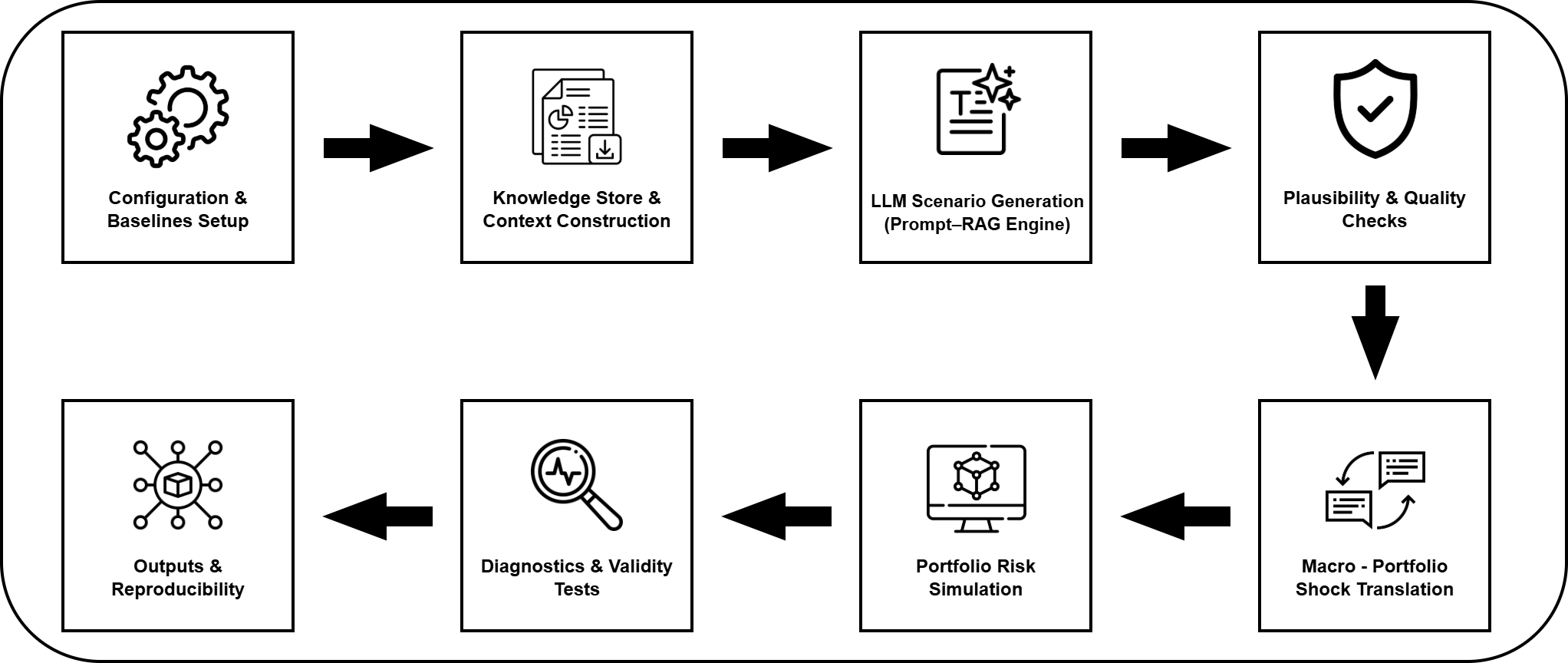}
  \caption{Pipeline for scenario generation and risk translation.  
  IMF fundamentals and optional news are embedded in MiniLM \cite{wang2020minilm} and retrieved via FAISS \cite{douze2025faiss}
  to condition the prompt. LLMs output structured JSON scenarios (GDP, inflation, interest rates, rationale,
  and sector-level exposures), which are screened by hard and soft plausibility gates
  and tagged with a regime label.  
  Scenario shocks are mapped to asset returns through three channels:
  (i) a pure volatility channel that scales the covariance matrix,
  (ii) a linear PCA factor channel, and
  (iii) a nonlinear polynomial factor channel that contains all text/RAG/news
  amplification.  
  Regime severity $\lambda$ mixes calm and stress-period covariance matrices, and all
  channels are benchmarked against deterministic, LLM-free baselines.
  Ablations toggle model type (GPT-5.6-luna vs.\ Llama-3.1-8B-Instruct),
  retrieval (RAG on/off), and news augmentation (on/off); the generation and
  risk-layer ablations are listed in Table~\ref{tab:ablations}.}
  \label{fig:pipeline}
\end{figure*}

We develop a retrieval-augmented LLM pipeline for generating structured macroeconomic
stress scenarios and translating them into portfolio tail-risk metrics.  
The system consists of five stages:  
(1) macro and market data ingestion,  
(2) document embedding and indexing for retrieval,  
(3) structured scenario generation by LLMs with plausibility filtering and regime tagging,  
(4) construction of deterministic, LLM-free baselines and regime-specific covariance
matrices, and  
(5) portfolio-level VaR/CVaR evaluation via three stress channels
(pure volatility, linear factor, and nonlinear factor).  
Figure~\ref{fig:pipeline} presents an overview.
Market factor extraction uses standard PCA methodology \cite{JolliffeCadima2016PCA, ConnorKorajczyk1986APT}, while historical volatility baselines follow RiskMetrics-style EWMA estimation \cite{JPMorgan1996RiskMetrics} and GARCH(1,1)–t models \cite{Engle1982ARCH, Bollerslev1986GARCH}.

\subsection{Pipeline Pseudocode (simplified)}

\begin{lstlisting}
# 1. Macro and news profiles
for country in G7:
    base = load_IMF_WEO(country)
    news = fetch_recent_news(country) if USE_NEWS else []
    profile = build_profile(base, news)
    embed = MiniLM(profile)
    faiss.add(country, embed)

# 2. Market data, PCA factors, and baselines (LLM-free)
prices = load_cached_prices(assets=ETF_UNIVERSE)   # SPY, IEF, GLD, sectors
pca_factors = fit_PCA_factors(prices[BALANCED_UNIVERSE],   # SPY,IEF,GLD,XLE,XLF,XLU
                              seed=SEED, standardise=True, sign_align=True)
betas_linear, betas_poly = estimate_factor_betas(prices, pca_factors)

Sigma_calm   = estimate_covariance(prices, period="2012-2019")
Sigma_crisis = estimate_covariance(prices, period=STRESS_WINDOWS)  # see Sec. 4.3
baselines = build_deterministic_baselines(WEO, prices, pca_factors)

# 3. Scenario generation across configurations
for cfg in grid(model ∈ {GPT-5.6-luna, Llama-3.1-8B-Instruct},
                RAG ∈ {0,1}, NEWS ∈ {0,1}):
    for prompt_variant in PROMPTS:       # 30 prompt variants
        ctx = retrieve(country, k=3) if RAG else [profile]
        out = LLM_generate(model, ctx, template=HORIZON)   # "Q4-2027"
        json = extract_JSON(out)

        if not passes_hard_plausibility(json):
            continue

        regime = NLI_classify_regime(json.rationale)
        lambda_severity = build_lambda(json, regime_score=regime.score)

        if not passes_soft_plausibility(json, lambda_severity):
            continue

        scenarios.append((json, lambda_severity, cfg))

# 4. Scenario-specific covariance and three stress channels
for s, lambda_severity, cfg in scenarios:
    Sigma_scen = (1 - lambda_severity)*Sigma_calm + \
                 lambda_severity*Sigma_crisis

    shocks_macro = macro_shock_vector(s)
    shocks_factors = macro_to_PCA(shocks_macro)

    mu_linear = linear_drift(betas_linear, shocks_factors, decay=True)
    mu_nonlinear = nonlinear_drift(betas_poly, shocks_factors,
                                   lambda_severity, cfg)

    # volatility-only channel: zero mean, scaled Sigma_scen
    paths_vol = simulate_paths(mu=0, Sigma=scale_cov(Sigma_scen, shocks_macro))

    # linear channel: mu_linear, Sigma_scen
    paths_lin = simulate_paths(mu=mu_linear, Sigma=Sigma_scen)

    # nonlinear channel: mu_linear + mu_nonlinear, Sigma_scen
    paths_nonlin = simulate_paths(mu=mu_linear+mu_nonlinear, Sigma=Sigma_scen)

    compute_VaR_CVaR_MDD(s, cfg, paths_vol, paths_lin, paths_nonlin, baselines)
\end{lstlisting}

\vspace{0.5em}

\subsection{Macroeconomic Inputs}
\label{sec:macro-inputs}

Country-level fundamentals come from the April~2026 \emph{IMF World Economic
Outlook} (WEO) \cite{IMF2026WEOdata}. For each G7 economy we extract the 2027
projections for real GDP growth and headline consumer-price inflation, which are
the baseline against which the model's Q4--2027 scenario is expressed. All seven
baselines are projections rather than outturns: the WEO records 2025 as the last
actual year for every G7 economy, and we flag this per country in the artefact
manifest.

The parse is validated against the IMF's own published G7 aggregate. Weighting
the seven country projections by purchasing-power-parity GDP reproduces the
published aggregate to within $0.001$~pp on both growth and inflation, which
guards against the silent failure modes---a shifted year column, a renamed
country, a mismatched indicator code---that a schema revision can introduce
without any visible symptom. The run aborts if the discrepancy exceeds
$0.10$~pp.

\paragraph{Policy-rate baseline.} The WEO carries no short-term policy rate.
Because the models emit a policy-rate \emph{level} for the horizon, a rate
baseline is required to express that level as a shock. We therefore supply the
current operational target of each G7 central bank as of 20~August~2026,
recorded with its source, decision date and vote where applicable: the Federal
Reserve at $3.625\%$ (target-range midpoint), the Bank of England at $3.75\%$,
the Bank of Canada at $2.25\%$, the European Central Bank deposit facility at
$2.25\%$ for Germany, France and Italy, and the Bank of Japan at $1.00\%$. The
Federal Reserve figure uses the target-range midpoint rather than the upper
bound, following the convention of cross-country policy-rate statistics.

Two consequences follow and are stated rather than assumed. First, the rate
baseline is held constant to the horizon---a random-walk assumption, standard in
stress testing but not innocuous, since a scenario that merely tracks market
pricing will register as tightening or easing. Forward-implied rates were
considered and rejected to avoid an additional data dependency; eliciting the
rate change directly is a candidate for future work. Second, Germany, France and
Italy share the ECB rate by construction, so any cross-country variation in the
rate channel among the three reflects the model's level choices rather than
baseline differences.

\paragraph{A frequency and vintage mismatch we do not fully resolve.} The three
components of the deviation vector are not defined on a common basis. Growth and
inflation deviations are measured against an \emph{annual} 2027 WEO projection
while the scenario is described as Q4--2027, and the policy-rate deviation is
measured against a \emph{spot} August-2026 rate held constant over five
quarters. A coherent contemporaneous vector would require either a
quarterly-frequency baseline for all three variables or an explicitly modelled
policy path. We can bound the consequence for the rate component: over the four
quarters preceding the run the G7 policy rates moved by a median of $0.25$~pp
per quarter, so a five-quarter random walk plausibly misstates the rate
deviation by of order $1$~pp, which is comparable to the interquartile range of
the emitted rate deviations themselves. Shifting every rate baseline by
$\pm 1$~pp (ablations S4 and S5) moves the pooled linear-channel VaR multiple
from $4.06\times$ to $4.31\times$ ($-1$~pp) and $3.81\times$ ($+1$~pp), about
$\pm 6\%$, while the volatility channel moves by about $1\%$: the
misspecification matters for the level of the linear-channel multiple but not
for any comparative conclusion. Growth and inflation are less affected
because a Q4 level and an annual average differ by less than the scenario-to-
scenario spread. We regard this as the most serious unresolved specification
issue in the paper and flag it accordingly rather than treating it as a minor
caveat.

\paragraph{Levels versus deviations.} The models are instructed to report
macroeconomic \emph{levels} for the horizon and, separately, the implied
\emph{change} versus the stated baseline. All downstream computation---the
factor mapping, the severity index, and every reported table---operates on
deviations, computed as level minus baseline. Reporting a policy-rate level of
$3.6\%$ as though it were a $3.6$~pp tightening inflates severity and
manufactures cross-country differences that are in fact baseline differences;
requiring both quantities allows the model's own arithmetic to be checked
(Section~\ref{sec:results-grounding}).

In addition to LLM-generated scenarios we construct deterministic, LLM-free
macro benchmarks: a fixed 2008--09-style global risk-off shock applied uniformly
($\Delta \text{GDP} = -3$~pp, $\Delta \text{inflation} = +1$~pp,
$\Delta \text{policy rate} = +1$~pp), and a uniform severe adjustment
($-5$, $+2$, $+1.5$~pp) applied to country-specific WEO projections. These bound
what the LLM contributes: any severity it produces must be read against them.

\subsection{Optional News Integration}
To capture high-frequency macro developments not yet reflected in the WEO,
the system optionally augments each country profile with recent
English-language news headlines. For each G7 country we query a news API
over a fixed lookback window anchored at the run timestamp, using a
country-specific boolean query on macro terms (e.g.\ “economy”, “GDP”,
“inflation”). If the provider rejects the requested window as being “too far
in the past”, the earliest admissible date is parsed from the error message
and the window is clamped forward to the maximum span allowed by the plan.

All retrieved articles are normalised to a fixed-size headline snapshot:
we sort deterministically by \texttt{publishedAt} and title, deduplicate on
title, and then truncate or pad to \textbf{exactly 50 rows per country}. When
fewer than 50 real headlines are available, the remaining rows are filled
with explicit \texttt{[PAD-XX] No headline available} markers so that every
\texttt{\_headlines.csv} file has the same schema and row count. These
50-row snapshots are saved to disk together with a JSON sidecar containing
the effective time window, query string, and retrieval attempts, and their
SHA-256 hashes are recorded in the run manifest.

For prompt construction we do not pass all 50 headlines directly. Instead,
we embed the real (non-padded) titles with MiniLM-L6-v2 and run
$k$-means clustering with $k=20$, selecting one exemplar per cluster. This
yields up to \textbf{20 diverse headlines per country}, which are inserted
as a “Top-20 diverse headlines” block in the context. If fewer than
20 real headlines exist, we simply return all of them. Padded rows are
never used in the prompt.

In this study, news headlines are \emph{snapshot-pinned}: the 50-row
headline CSV and its JSON metadata are fetched once per country, written
to disk, and treated as immutable. All reported results are computed using
these frozen headline snapshots; subsequent analyses never re-query the
news API.

\subsection{Knowledge Store Construction}

Each country profile (baseline WEO data, with or without news) is serialized to  
plain text and embedded using MiniLM-L6-v2.  
Embeddings are indexed with FAISS using flat 
inner-product search. The retrieval index is deterministic: the input corpus, ordering, and random seed are fixed so that top-$k$ neighbours are reproducible.  
Raw documents are hashed and cached for auditability. To guarantee replayability, all inputs to retrieval are snapshot-frozen. This includes: (i) serialized WEO baseline files, (ii) headline CSVs, (iii) MiniLM embeddings, and (iv) the FAISS index itself. All files are hashed and recorded in a run manifest. Retrieval is therefore deterministic conditional on these frozen artefacts.

\subsection{Semantic Retrieval}

For a given query country, the retriever returns the top-$k = 3$ most relevant
country profiles by cosine similarity.  
These tend to be economically similar peers (e.g.\ U.S.\ retrieves Canada and the U.K.)
and provide contextual anchors for the LLM.  
If RAG is disabled, the prompt includes only the target country's profile.
Retrieval depth $k$ balances contextual richness with token-length constraints.

\subsection{Prompt Construction}

Prompts consist of  
(i) a system instruction positioning the LLM as a macro–financial analyst,  
(ii) a context block with retrieved WEO fundamentals (and optionally news), and  
(iii) a directive to generate a severe but internally coherent macroeconomic scenario for Q4~2027.

We use ``counterfactual'' throughout in the ordinary sense of a constructed
alternative to the baseline projection, as the stress-testing literature does,
and not in the causal-inference sense: no intervention is identified and no
potential-outcome claim is made.

The model must return a structured JSON object with:
\texttt{country}, \texttt{title},  
\texttt{gdp\_growth}, \texttt{inflation}, \texttt{interest\_rate},  
\texttt{rationale}, and \texttt{risk\_sectors}.  
A tolerant parser extracts the first valid JSON object.

\subsection{Language Model Inference}

We evaluate two compact instruction–tuned models:

\begin{itemize}
  \item \textbf{GPT-5.6-luna} (OpenAI, 2026),
  \item \textbf{Llama-3.1-8B-Instruct} (Meta, 2024).
\end{itemize}

The full experimental grid (countries, model family, retrieval flag, news flag, and
prompt variants) is described in Section~\ref{sec:experimental}. For each
(country, model, retrieval, news) configuration we generate \emph{30 distinct
prompt variants}, and we produce \emph{one scenario per prompt variant}. Thus each
configuration yields exactly 30 raw scenarios before plausibility filtering.

\textbf{Determinism note.}
Decode parameters are recorded as transmitted rather than as configured. The
GPT-5 family accepts only $\text{temperature}=1$ and does not support greedy
decoding, so strict determinism is unavailable at the API level
\cite{staudinger2024reproducibility}; a reasoning-effort setting of
\texttt{none} is used and logged. We therefore describe the two GPT runs as
independent repeat draws under identical settings, not as a temperature
ablation, and quantify their agreement in Section~\ref{sec:results-reproducibility}.
Residual variation arises from provider-side batching and kernel selection, and
from retrieval ordering when cosine similarities are near-tied. Reproducibility
claims in this paper are conditional on snapshot-frozen context---the embedded
WEO baseline, MiniLM embeddings pinned to CPU execution, the FAISS index, and a
hash-recorded news headline snapshot---and are made at the level of aggregate
statistics rather than individual scenarios.

\textbf{Plausibility audit and regime tagging.}
\label{sec:plausibility}
Each generated scenario is screened by a two-layer filter. A hard gate operates
on the emitted levels and rejects implausible or internally contradictory macro
values ($|\text{GDP growth}| > 10\%$, inflation $> 20\%$, policy rate outside
$[-1\%, 15\%]$, or a deep recession paired with disinflation and rate hikes
absent a credible rationale) \cite{hardy2014stress}. A scenario is also
rejected if the model's own stated changes are missing or exceed the same bounds
in deviation space ($|\Delta g| > 10$, $|\Delta\pi| > 20$, $|\Delta r| > 15$~pp). A minimum growth-shortfall
requirement is applied to the \emph{deviation} rather than the level, since a
level-based floor would impose a different effective severity on each country
and truncate the reported distribution
\cite{hopper2022designing}. A soft score in $[0,5]$ then rewards rationale
substance and cross-variable coherence, with an acceptance threshold of $2.5$.
The deviation floor is $\Delta g_s \leq -0.5$~pp; it binds for the
open-weight model, whose least severe accepted scenario sits at $-0.503$~pp.

We use ``schema-valid and internally consistent'' rather than ``plausible''
wherever external validity has not been demonstrated. A rule-based gate that
rejects $0.2\%$ of one model's output establishes that the output is
well-formed and not self-contradictory; it does not establish that a
macroeconomist would judge the scenarios realistic, and we make no such claim.

In parallel we classify the free-text rationale with a DeBERTa NLI model
\cite{he2021debertav3} into \emph{normal}, \emph{stress} or \emph{crisis},
obtaining a continuous regime score
$\rho_s = 0\cdot p_{\text{normal}} + 0.5\cdot p_{\text{stress}} + 1\cdot p_{\text{crisis}}
\in [0,1]$.

\paragraph{Severity index.} The macro deviations and the regime score combine
into the scalar severity index used for covariance mixing, with equal weight on a
macro component and on the narrative score:
\begin{equation}
\begin{aligned}
\lambda_s &= \tfrac{1}{2}\,\lambda^{\text{macro}}_s + \tfrac{1}{2}\,\rho_s,\\
\lambda^{\text{macro}}_s &= \bigl(0.5\,u_{g,s} + 0.3\,u_{\pi,s} + 0.2\,u_{r,s}\bigr)^{1.3},\\
u_{g,s} &= \operatorname{clip}_{[0,1]}\!\bigl((-\Delta g_s - 0.5)/4.5\bigr),\\
u_{\pi,s} &= \operatorname{clip}_{[0,1]}\!\bigl((\Delta \pi_s - 2)/6\bigr),\\
u_{r,s} &= \operatorname{clip}_{[0,1]}\!\bigl(\Delta r_s/5\bigr),
\end{aligned}
\label{eq:lambda}
\end{equation}
so the growth score is zero at the $-0.5$~pp deviation floor and saturates at a
$5$~pp shortfall, the inflation score is zero below a $+2$~pp surprise and
saturates at $+8$~pp, and the rate score saturates at a $5$~pp tightening;
easing and disinflation contribute nothing. These weights, breakpoints and the
convexity exponent are set a priori and not tuned; we report the realised distribution of $\lambda_s$ in
Section~\ref{sec:results-risk} so the reader can judge how much of the
covariance mixture is driven by narrative versus macro content.

\paragraph{Volatility scaling.} The pure volatility channel scales the
scenario covariance by a deterministic function of the inflation deviation,
\begin{equation}
\Sigma^{\text{vol}}_{\text{scen},s} = c_s^2\,\Sigma_{\text{scen},s},
\qquad
c_s = \max\bigl(0.1,\; 1 + \max(0,\Delta \pi_s)/20\bigr),
\label{eq:volscale}
\end{equation}
so that a $+4$~pp inflation surprise raises the volatility scale by $20\%$.

\paragraph{Drift decay and amplification coefficients.} The shocked drift decays
geometrically at $\delta = 0.98$ per trading day, giving a half-life of
approximately 34 days and a cumulative multiplier of $36.0$ over the 63-day
horizon; the value is chosen so that roughly two thirds of the shock has
dissipated by the end of the quarter, consistent with quarter-ahead shock
reversion in supervisory practice. The amplification coefficients in the
nonlinear channel are hand-set at $a_\lambda = 0.30$ and $a_{\text{rag}} =
a_{\text{news}} = a_{\text{rag}\times\text{news}} = 0.03$. They are \emph{not} fitted, cross-validated or taken
from prior work, and we flag them as an undisclosed calibration layer in earlier
formulations of this pipeline. Section~\ref{sec:results-risk} shows that the
channel they scale contributes less than $0.6\%$ of the VaR or CVaR estimate, so the reported
results are insensitive to them; that insensitivity is the reason we did not
attempt to estimate them.

Two automated cross-checks supplement the gate and proved more diagnostic than
the gate itself. The first reconciles the model's stated change against the
change implied by its reported level and baseline; the second reports the
dispersion of the growth shock and warns when it collapses onto the prompted
severity floor. Both surfaced failures invisible to the plausibility rules and
to a reader of the narratives (Section~\ref{sec:results-tiers} and
Section~\ref{sec:disc-models}). We report the rejection rate explicitly for
every run rather than describing the gate as a filter: on the primary run
it rejects $0.2\%$ of scenarios, and a gate that rejects almost nothing should
be characterised as a guardrail, not a screen.

\subsection{Ablation Dimensions}

Throughout the paper we vary three binary design choices in addition to country:

\begin{itemize}
  \item \textbf{Model family} (GPT-5.6-luna vs.\ Llama-3.1-8B-Instruct),
  \item \textbf{Retrieval (RAG)} enabled vs.\ disabled,
  \item \textbf{News retrieval} enabled vs.\ disabled.
\end{itemize}

Taken together, these switches define the eight system configurations per country that we refer to simply as “configurations” in Sections~\ref{sec:experimental}–\ref{sec:results}, where we study their effects on plausibility, dispersion, tail risk, ANOVA variance decomposition, and cross-country consistency diagnostics.

On top of this grid, every experimental variation is declared as a named preset
of seven axes (Table~\ref{tab:ablations}). Two axes change the \emph{generated
scenarios}---generation mode and prompt template set---and key the scenario
cache; the remaining five change only the \emph{risk layer}---the macro--equity
calibration $\kappa$, the factor-extraction universe, the shock source, the
stress-covariance estimator and a shift of the policy-rate baseline---and reuse
the cached scenarios of their generation without further API calls. Each preset
writes to its own results folder, so no two conditions can overwrite each other.
Presets P1--P4 are generation conditions; S1--S8 vary one axis of P1 (or of P3
for S8) at a time.

\begin{table*}[t]
  \centering
  \small
  \caption{Ablation presets. Each row changes the listed axis relative to the
  primary run P1 (GPT-5.6-luna, tiered severity bands, $\kappa = 3.5\%$,
  balanced six-asset PCA universe, computed shocks, blended stress covariance,
  unshifted rate baseline).}
  \label{tab:ablations}
  \begin{tabular}{lll}
    \toprule
    Preset & Scope & Change relative to P1 \\
    \midrule
    P1 & --- & Primary run (headline results) \\
    P2 & generation & Independent repeat draw of P1 under identical settings \\
    P3 & generation & Llama-3.1-8B-Instruct instead of GPT-5.6-luna \\
    P4 & generation & Unanchored templates: no growth-shortfall band \\
    S1 / S2 & risk layer & $\kappa = 2.0\%$ / $5.0\%$ \\
    S3 & risk layer & $\kappa$ estimated empirically (Section~\ref{sec:calibration}) \\
    S4 / S5 & risk layer & Policy-rate baseline shifted by $-1$ / $+1$~pp \\
    S6 & risk layer & Complete-case stress covariance only (no GFC observations) \\
    S7 & risk layer & PCA extracted from SPY, IEF, GLD only (degenerate) \\
    S8 & risk layer & P3 with the model's \emph{stated} changes as shocks \\
    \bottomrule
  \end{tabular}
\end{table*}

\subsection{Portfolio Stress Mapping (Three-Channel Factor Model)}

Scenario shocks are mapped into portfolio losses using a  
\textbf{three-factor PCA model} on ETF returns \cite{FamaFrench1993Factors}, together with
calm/stress covariance mixtures and three distinct stress channels.

\paragraph{Portfolios.}
We consider two representative portfolios:

\begin{itemize}
  \item \textbf{Portfolio A:} U.S.\ equity (SPY), intermediate Treasuries (IEF),
        and gold (GLD) with fixed weights of 60/30/10.
  \item \textbf{Portfolio B:} an equal-weighted portfolio across the full set of
        sector ETFs with sufficient history (XLE, XLF, XLK, XLY, XLI, XLU,
        XLV, XLP, XLB, XLRE).
\end{itemize}

Weights are re-normalized daily over a 63-day horizon. We use a 63-day horizon because it corresponds to approximately one trading
quarter, matching the Q4~2027 shock horizon specified in the scenario prompts.
Quarter-ahead propagation is standard in macro–financial stress testing
(e.g., CCAR, ECB, BoE) and provides a stable yet responsive window for PCA
factor estimation: shorter windows (e.g., 21 days) produce noisy loadings,
while longer windows (e.g., 126–252 days) dilute the impact of the intended
macro shock by averaging over overly long historical periods.

\paragraph{PCA factors and betas.}
We extract three principal components from the daily excess returns of a
balanced six-asset universe---SPY, IEF, GLD, XLE, XLF and XLU---standardised
before extraction so that correlated equity sectors do not crowd out the
duration and safe-haven dimensions. Extraction from the three portfolio assets
alone is degenerate: three components span three assets exactly, every curvature
term fits to machine precision, and the nonlinear channel is identically zero
for Portfolio~A. Extraction from the full thirteen-asset universe leaves gold
poorly identified because eleven of thirteen assets are equity-like.

Components are assigned to economic roles by solving an assignment problem on
absolute correlation with three anchors (SPY for equity, GLD for safe-haven,
IEF for duration) rather than by variance rank, which would mislabel the factors
whenever the variance ordering differs from the economic ordering, and each is
then sign-aligned to load positively on its anchor. Under the balanced universe
the resulting anchor correlations are $0.951$ (PC1--SPY), $0.582$ (PC2--GLD) and
$0.833$ (PC3--IEF), with explained variance shares of $0.464$, $0.239$ and
$0.116$. We omit the full anchor-correlation matrix for space and
describe PC2 as a gold-tilted safe-haven factor rather than a pure gold factor,
since PC3 also loads materially on gold.

Betas are estimated on standardised factors with an intercept, so a beta reads
as return per one standard deviation of factor and is dimensionally consistent
with the standardised shocks applied downstream. Nonlinear betas add squares and
cross-terms of the three factors.

\paragraph{Calibration of the macro-to-drift scale.}
\label{sec:calibration}
The constant $\kappa$ converting factor-standard-deviation units into a daily
drift sets the level of every reported multiple. Rather than fix it by hand, we
calibrate it so that a one percentage point shortfall in GDP growth moves the
equity anchor by a stated fraction over the horizon, and we report results
across $\kappa \in \{2\%, 3.5\%, 5\%\}$
(Section~\ref{sec:results-sensitivity}). A drift cap is retained as a numerical
safeguard but is derived from the largest shock the plausibility gate admits (the
drift implied by a $12$~pp growth shortfall), which lies beyond the largest growth shortfall accepted in any run
($10.6$~pp), so the cap is a safeguard against malformed inputs rather than a
censor of the severe tail.

We also attempted to replace the assumed $\kappa$ with an estimate (ablation
S3). Using 43 non-overlapping quarters (2015Q4--2026Q2) of US real GDP growth,
CPI inflation and the effective federal funds rate, the association between
quarterly macro changes and contemporaneous cumulative factor returns is
significant only for the equity factor, and with the wrong sign
($-0.19$ factor standard deviations per percentage point, HC1 standard error
$0.08$, $p = 0.02$, $R^2 = 0.19$); the inflation and duration elasticities are
insignificant ($p = 0.12$ and $0.22$). The implied $\kappa$ has a signed $95\%$
interval of $[-0.24\%, +0.12\%]$, which contains zero and excludes every value
in our range. The empirical calibration is therefore unusable, the S3 run falls
back to $\kappa = 3.5\%$ and reproduces P1 exactly, and we report it as a
negative result rather than as a fourth calibration.

LLM macro shocks are given in percentage points as
$\Delta \mathbf{M}_s = (\Delta g_s, \Delta \pi_s, \Delta r_s)$, each measured as
a deviation from the baseline of Section~\ref{sec:macro-inputs}.

To connect these to the PCA factors \cite{oura2012macrofinancial} we define the
signed factor shock vector in Equation~\ref{eq:factor-shock}. The signs are
oriented so that an adverse growth shock loads negatively on the equity factor
and a rate rise loads negatively on the duration factor, which is the
arrangement under which a deeper recession produces a larger loss. We retain the
sign rather than truncating at zero: truncation discards every disinflationary
and easing scenario, and in our corpus inflation and rate deviations are
two-sided in roughly a third of cases.

\begin{equation}
\label{eq:factor-shock}
\Delta \mathbf{F}_s=\bigl(f_s^{(1)},f_s^{(2)},f_s^{(3)}\bigr)
=\Bigl(
      \Delta g_s/100,\;
      \Delta \pi_s/100,\;
      -\Delta r_s/100
\Bigr).
\end{equation}

\paragraph{Regime-specific covariance.}
From historical ETF returns we estimate two covariance matrices:
$\Sigma_{\text{calm}}$ using a calm sample (2012--2019, 1{,}064 complete-case
days, Ledoit--Wolf shrinkage) and
$\Sigma_{\text{stress}}$ estimated on selected stress windows (Section~\ref{sec:experimental}).  
For each scenario $s$, the regime severity index $\lambda_s \in [0,1]$
constructed in the plausibility step defines a scenario-specific covariance:
\begin{equation}
\Sigma_{\text{scen},s}
  = (1 - \lambda_s)\,\Sigma_{\text{calm}} + \lambda_s\,\Sigma_{\text{stress}}
\label{eq:cov-mix}
\end{equation}
Higher $\lambda_s$ values thus induce more crisis-like volatilities and
correlations while preserving a purely linear, interpretable mixture between
the two regimes \cite{gray1996modeling, HamiltonLin1996Volatility}.

\paragraph{Stress channels and simulation.}
We decompose stress transmission into three channels:

\begin{enumerate}
  \item \textbf{Pure volatility channel.}  
  The covariance matrix $\Sigma_{\text{scen},s}$ is scaled as a deterministic
  function of the inflation shock (larger inflation $\Rightarrow$ higher
  volatility), while daily mean returns are kept at zero. This isolates the
  impact of volatility-only amplification on VaR/CVaR.

  \item \textbf{Linear PCA factor channel.}  
  Macroeconomic shocks are mapped to PCA factor shocks $\Delta \mathbf{F}_s$,
  which are expressed in units of the daily volatility $\sigma^{\text{anc}}_j$
  of each factor's anchor asset (SPY, GLD, IEF), passed through the linear
  betas and scaled by the calibrated constant $k(\kappa)$ to obtain a daily
  drift shift for each asset $i$,
  $\mu_{\text{lin},s,i} = \operatorname{clip}_{[-\bar\mu,\bar\mu]}
  \bigl(k(\kappa)\sum_j \beta_{ij} f^{(j)}_s/\sigma^{\text{anc}}_j / 63\bigr)$,
  where $\bar\mu$ is the drift cap. No baseline drift is added, so the channel
  isolates the scenario-induced shift.
  The resulting drift is applied with a geometric decay over the 63-day horizon
  to reflect shock reversion.  
  Crucially, this channel is independent of text, retrieval, news, and $\lambda$:
  it depends only on the numeric macro shocks and the pre-estimated PCA mapping.

  \item \textbf{Nonlinear factor channel.}
  The curvature terms of the polynomial expansion---squares and cross-terms
  only---are applied to the same factor shocks to produce an \emph{increment}
  over the linear drift \cite{Kutateladze2022KernelTrick, almeida2023nonlinear},
  which is then multiplied by an amplification term of the form
  $\text{amp}_s = 1 + a_\lambda \lambda_s + a_{\text{rag}}\mathbf{1}_{\text{RAG}}
  + a_{\text{news}}\mathbf{1}_{\text{NEWS}}
  + a_{\text{rag}\times\text{news}}\mathbf{1}_{\text{RAG}}\mathbf{1}_{\text{NEWS}}$
  with small fixed coefficients. Including the linear terms of the expansion
  here would double-count the linear channel and yield a merely rescaled copy of
  it rather than an increment. Per-asset caps ensure numerical stability. This
  is the \emph{only} channel through which text, retrieval and news affect
  portfolio returns, and Section~\ref{sec:results-risk} shows that in practice
  it contributes less than $0.6\%$ of the VaR or CVaR estimate---a design limitation we report
  rather than an amplification effect we claim.
\end{enumerate}

Daily returns are simulated using a Cholesky decomposition of
$\Sigma_{\text{scen},s}$ (with jitter if needed), geometric drift decay, and
per-day clipping of extreme returns \cite{jorion1997value, glasserman2004monte, Best2000VaR}. For the volatility channel we set the mean
to zero and use the scaled $\Sigma_{\text{scen},s}$; for the linear and nonlinear
channels we use $\Sigma_{\text{scen},s}$ together with $\mu_{\text{lin},s}$ and
$\mu_{\text{lin},s} + \mu_{\text{nonlin},s}$, respectively.  
In each case we simulate 20,000 paths of daily returns over a 63-day horizon.

\paragraph{Tail metrics.}
For each simulated path we compute portfolio-level losses and derive:

\begin{itemize}
  \item 5\% Value-at-Risk (VaR\(_{0.95}\))~\cite{jorion1997value},
  \item Conditional VaR (CVaR)~\cite{rockafellar2000optimization},
  \item Maximum drawdown (MDD)~\cite{magdon2004maximum}.
\end{itemize}

These metrics are computed separately for the volatility-only, linear, and
nonlinear channels.  
Severity is expressed as the ratio relative to the historical bootstrap or
econometric baseline (i.e., VaR/CVaR \emph{multiples}).  
These multiples underpin the risk tables and figures reported in
Section~\ref{sec:results}, and allow us to compare LLM-induced stress against both deterministic macro benchmarks and classical volatility models (EWMA and GARCH-t).

\section{Experimental Setup}
\label{sec:experimental}

This section describes the experimental design used to evaluate the LLM-based
stress–testing pipeline.  
Macroeconomic stress scenarios are generated for the G7 under the configuration
grid defined by the ablations in Section~\ref{sec:methodology}, filtered via
plausibility checks and regime tagging, and then mapped into portfolio tail–risk
metrics through the three stress channels described earlier.    
The resulting VaR and CVaR multiples (volatility, linear, and nonlinear) are
benchmarked against historical and econometric baselines and later summarised in
the tables and figures reported in Section~\ref{sec:results}.

\subsection{Language Models and Runs}

We evaluate two instruction-tuned large language models:

\begin{itemize}
  \item \textbf{GPT-5.6-luna} (proprietary, OpenAI), used as the main workhorse model.
  \item \textbf{Llama-3.1-8B-Instruct} (open source, Hugging Face), used for cross--model comparisons.
\end{itemize}

Completions are required to emit a single JSON object; we parse the first valid object with a tolerant extractor \cite{Wu2024StructuredEntity} and discard malformed outputs (rare, $<2\%$).

\paragraph{Model identification.} GPT-5.6-luna was accessed through the OpenAI
Chat Completions API between 20 and 22 August 2026 using the snapshot string
\texttt{gpt-5.6-luna}; the decode configuration transmitted with every request
was \texttt{max\_completion\_tokens}~$=1200$, \texttt{seed}~$=42$ and
\texttt{reasoning\_effort}~$=$~\texttt{none}, with temperature left at the
provider default of $1$ because the GPT-5 family accepts no other value.
Llama-3.1-8B-Instruct (released 2024) was accessed through the Hugging Face
Inference Providers routing layer at revision
\texttt{5206a32e0bd3067aef1ce90f5528ade7d866253f} with
\texttt{temperature}~$=0.9$, \texttt{top\_p}~$=0.9$ and
\texttt{max\_tokens}~$=900$. All 30 prompt templates are reproduced verbatim in
Appendix~\ref{app:prompts}.

\paragraph{Model and hardware details.}
GPT-5.6-luna is accessed through the OpenAI API and 
Llama–3.1–8B–Instruct via the Hugging Face 
Inference API. We do not apply any custom quantisation, fine–tuning, 
or weight modifications: both models are used exactly as provided by their 
respective hosted endpoints. We rely on the providers’ default inference 
settings (including their default sampling parameters and seed behaviour) and 
do not enforce strict greedy decoding. All prompts, including retrieved 
context, are constructed to remain well within the provider–exposed context 
windows, and we therefore do not impose any additional truncation rules 
beyond the usual safety checks for malformed JSON.

For retrieval\textendash augmented generation (RAG), we deliberately fix the 
retrieval depth at top–$k=3$ documents. This choice reflects a balance between 
contextual richness, latency, and prompt readability, rather than any hard 
capacity constraint imposed by the models or infrastructure. Inference runs on 
provider\textendash hosted CPU endpoints (no GPU acceleration), but the pipeline 
itself is hardware\textendash agnostic, and all experiments operate comfortably 
within the resource limits of standard hosted inference services.

We run three complementary experiments:

\begin{enumerate}
  \item A \emph{primary GPT-5.6-luna run} (\texttt{run\_A}, preset P1), which forms the core dataset for the macro, risk, stability and consistency results in Section~\ref{sec:results}.
  \item A \emph{repeat GPT-5.6-luna run} (\texttt{run\_B}, preset P2): an independent redraw of every cell with byte-identical request parameters, retrieval artefacts and news snapshot. The two GPT runs differ only in the draw; the primary run is deduplicated on its input hashes and the repeat run on its output hashes, which affects bookkeeping but not the requests sent.
  \item A \emph{Llama-3.1-8B-Instruct run} (\texttt{run\_llama}, preset P3) used for cross--model severity and risk comparisons, reported alongside the deterministic baseline in Section~\ref{sec:results}.
\end{enumerate}

Across the three runs we obtain 838 (primary GPT-5.6-luna run), 838 (repeat
GPT-5.6-luna run) and 744 (Llama-3.1-8B-Instruct) accepted scenarios after the
two-layer plausibility audit of Section~\ref{sec:plausibility}, from 840
intended scenarios per run. Rejection rates are $0.2\%$, $0.2\%$ and $11.4\%$
respectively. A fourth generation (preset P4) repeats the primary run with
templates that prescribe no growth-shortfall band; all 840 of its
scenarios are accepted. All three runs use an identical prompt template set (the tiered set, hash
\texttt{1d7e6dee3dc7}; the unanchored set used by P4 hashes to
\texttt{02005231eafd}), recorded by hash in each scenario record so that cross-model comparisons cannot silently
conflate model behaviour with template revision. These accepted scenarios form
the inputs to the volatility, linear and nonlinear stress channels.

\paragraph{Run attribution.}
Unless explicitly stated otherwise, all tables and figures in the main text are based on the primary GPT-5.6-luna run (\texttt{run\_A}). Results that use the non-primary GPT-5.6-luna run (\texttt{run\_B}) or the Llama-3.1-8B-Instruct run (\texttt{run\_llama}) are explicitly identified in the surrounding text or figure/table captions.

\subsection{Configuration Grid and Scenario Sampling}

The main primary run uses a full factorial grid over:

\begin{itemize}
  \item \textbf{Countries:} G7 (Canada, France, Germany, Italy, Japan, United Kingdom, United States).
  \item \textbf{Retrieval (RAG):} on vs.\ off.
  \item \textbf{News retrieval:} on vs.\ off.
  \item \textbf{Prompt variant:} 30 manually designed macro narratives per country.
\end{itemize}

For each (country, RAG, news) configuration we generate one scenario per prompt variant, yielding
\[
  7\ \text{countries} \times 4\ \text{configs} \times 30\ \text{prompts} = 840
\]
intended scenarios per model.  
After filtering out malformed outputs and scenarios failing the hard or soft
plausibility gates, the primary GPT-5.6-luna run yields 119--120 accepted
scenarios per country (Table~\ref{tab:macro-summary}, last column), for a total
of 838; the Llama run yields 91--117 per country, for a total of 744.

Scenario stability is quantified ex post using two dispersion metrics derived from these samples:

\begin{itemize}
    \item \textbf{Intra-prompt dispersion} (Table~\ref{tab:prompt-dispersion}): for each prompt variant and configuration, we pool the accepted scenarios for that prompt across runs and compute the average pairwise distance between macro-shock vectors
(gdp\_growth, inflation, interest\_rate).
  \item \textbf{Intra-configuration dispersion} (Table~\ref{tab:stability-config}): for each (country, RAG, news) cell in the primary run, we compute the average pairwise distance across all accepted scenarios in that cell, with bootstrap confidence intervals.
\end{itemize}

\paragraph{Dispersion metric.}
For a given cell (e.g., a fixed country, model, RAG, and news configuration)
with $S$ accepted scenarios, let
$\mathbf{x}_s = (\Delta g_s, \Delta \pi_s, \Delta r_s)$ denote the vector of
GDP, inflation, and interest-rate shocks (in percentage points) for scenario $s$.
We measure stability using the dispersion metric in
Equation~\ref{eq:dispersion}, a standard approach based on mean 
pairwise Euclidean distance, 
commonly used to assess diversity and robustness in generative models 
and scenario generators \cite{FreieslebenGrote2023Robustness}.

\begin{equation}
D = \frac{2}{S(S-1)} \sum_{1 \le i < j \le S}
  \left\lVert \mathbf{x}_i - \mathbf{x}_j \right\rVert_2.
\label{eq:dispersion}
\end{equation}

No additional scaling or standardisation is applied; all shocks are measured in
percentage points for comparability with
Tables~\ref{tab:macro-summary} and~\ref{tab:severity-model}.
This metric is used both at the prompt level (Table~\ref{tab:prompt-dispersion})
and at the configuration level (Table~\ref{tab:stability-config}).

One corrupted configuration and one corrupted prompt-level row with extremely large dispersion were removed via simple QC filters (threshold 20 in the shock space), as reported in the table logs.

\subsection{Portfolio and Econometric Baselines}

All scenarios are mapped to two stylised ETF portfolios:

\begin{itemize}
  \item \textbf{Portfolio A} (headline portfolio): 
    \begin{itemize}
      \item \textbf{SPY} — U.S.\ equities (S\&P~500 proxy) — 60\% weight,
      \item \textbf{IEF} — intermediate-term U.S.\ Treasuries — 30\% weight,
      \item \textbf{GLD} — gold bullion — 10\% weight.
    \end{itemize}
  \item \textbf{Portfolio B} (sector-tilted robustness portfolio): an equal-weighted portfolio across sector ETFs with full history (XLE, XLF, XLK, XLY, XLI, XLU, XLV, XLP, XLB, XLRE).
\end{itemize}

Weights are re-normalised daily over a 63-day horizon (one quarter) to maintain
constant-weight exposure.  
Portfolio~A is the main focus of the paper; Portfolio~B is used to test
cross--portfolio robustness in the cross-country consistency diagnostics
(Table~\ref{tab:fairness}).

To provide non-LLM risk baselines for Portfolio~A, we estimate 63-day VaR and
CVaR under three standard methods (Table~\ref{tab:baseline-var}; see also
Figure~\ref{fig:baselines} in Section~\ref{sec:results}).  
All three are constructed from overlapping 63-day windows of historical returns:

\begin{itemize}
  \item \textbf{Historical Baseline (Bootstrap)} on daily returns over the full common
        sample (November~2004 onward for Portfolio~A, constrained by the listing
        history of gold),
        sampling overlapping 63-day blocks with replacement \cite{efron1992bootstrap, politis1994stationary}.
  \item \textbf{EWMA} with decay factor $\lambda = 0.94$ under a Normal assumption, rescaled to the 63-day horizon.
  \item \textbf{GARCH(1,1)–t} fitted to daily returns, with 63-day losses obtained from simulated paths.
\end{itemize}

Table~\ref{tab:baseline-var} reports the resulting 63-day loss estimates in
decimal units for both portfolios.

\begin{table*}[t]
  \centering
  \caption{63-day VaR and CVaR under historical and econometric baselines. The
  bootstrap is available for both portfolios; the EWMA and GARCH benchmarks are
  estimated for Portfolio~A only.}
  \label{tab:baseline-var}
  \begin{adjustbox}{max width=\textwidth}\begin{tabular}{llcc}
    \toprule
    Portfolio & Method &
    \makecell{VaR$_{0.95}$\\(decimal loss)} &
    \makecell{CVaR$_{0.95}$\\(decimal loss)} \\
    \midrule
    A & Historical Baseline (Bootstrap) & 0.0490 & 0.0911 \\
    A & EWMA ($\lambda=0.94$, Normal)   & 0.0758 & 0.0952 \\
    A & GARCH(1,1)--$t$ (Simulated)     & 0.0793 & 0.1090 \\
    \midrule
    B & Historical Baseline (Bootstrap) & 0.0727 & 0.1384 \\
    \bottomrule
  \end{tabular}\end{adjustbox}
\end{table*}

We report multiples against all three benchmarks rather than against the
bootstrap alone. The choice of denominator matters materially: the same absolute
losses correspond to $3.94\times$ VaR against the bootstrap, $2.55\times$ against
EWMA and $2.43\times$ against GARCH$(1,1)$--$t$
(Table~\ref{tab:multi-baseline}). Reporting only against the smallest
denominator would flatter the result, and we avoid it. The bootstrap remains the
default reference because it is unconditional whereas EWMA and GARCH are
conditional on the run date, but none of the three is a like-for-like comparator
for a Gaussian covariance simulation: the bootstrap preserves empirical serial
dependence and tail shape, and GARCH$(1,1)$--$t$ is a different stochastic
process altogether. Any statement of the form ``LLM scenarios imply $N\times$
VaR'' is therefore conditional on the denominator, the covariance regime, the
distributional assumptions, the macro-to-drift mapping, $\kappa$ and the
shock-reversion specification. Note that risk does not increase
monotonically with model flexibility: the bootstrap CVaR exceeds the EWMA CVaR
because the historical sample is more heavily left-tailed than the Normal
assumption underlying the EWMA rescaling, even though the EWMA VaR is higher.  
Section~\ref{sec:crisis-envelopes} additionally reports crisis envelopes for
Portfolio~A by comparing GFC and COVID windows against both this unconditional
baseline and a calm-period (2012--2019) bootstrap; these envelopes are used only
for historical episode benchmarking and do not replace the main baseline in
Table~\ref{tab:baseline-var} for LLM-generated scenario multiples.

\paragraph{Stress-period covariance composition.} The regime-mixed covariance requires
a stress sample estimated on the full traded universe. We pool three windows:
the Global Financial Crisis (September~2007--June~2009, 460 trading days), the
COVID-19 crash (15~February--31~May~2020, 72 days) and the 2022 inflation shock
(January--October~2022, 209 days). Because the real-estate sector ETF lists only
from October~2015, a complete-case estimate across all thirteen assets retains
only 281 stress days and no GFC observations. We therefore blend, with equal
weight, (i) a pairwise estimate that uses every stress day on which both legs
trade (741 days in total, at least 281 per pair, with GFC history for 12 of the
13 assets), repaired to the nearest positive semi-definite matrix, and (ii) a
Ledoit--Wolf complete-case estimate. Replacing the blend by the complete-case
estimate alone (ablation S6) changes the pooled multiples by less than
$0.2\%$ (linear VaR $4.060\times$ against $4.058\times$; volatility VaR
$2.818\times$ against $2.816\times$), so the headline results do not depend on
whether the GFC enters the stress sample.

\subsection{Shock Translation and Tail-Risk Metrics}

For each accepted scenario $s$, the LLM produces macro shocks in percentage points for real GDP growth, inflation, and the interest rate.  
These shocks are mapped into portfolio tail risk via the three-channel factor
model described in Section~\ref{sec:methodology}.  
In brief, we estimate linear and nonlinear PCA factor betas for each asset,
construct a scenario-specific covariance matrix
$\Sigma_{\text{scen},s} = (1-\lambda_s)\Sigma_{\text{calm}} +
\lambda_s\Sigma_{\text{stress}}$, and simulate 63-day paths under:

\begin{itemize}
  \item a \emph{pure volatility} channel (zero drift, scaled covariance),
  \item a \emph{linear} channel (drift from the linear PCA mapping only),
  \item a \emph{nonlinear} channel (drift from linear + polynomial terms,
        amplified by $\lambda_s$ and the RAG/news flags).
\end{itemize}

Let $\text{VaR}^{\text{base}}$ and $\text{CVaR}^{\text{base}}$ denote the
63-day VaR and CVaR of Portfolio~A under the historical bootstrap baseline, and
let $\text{VaR}^{\text{(ch)}}_s$ and $\text{CVaR}^{\text{(ch)}}_s$ be the
corresponding quantities under scenario $s$ in a given channel
$\text{ch} \in \{\text{vol}, \text{lin}, \text{nonlin}\}$.  
We compute VaR and CVaR multiples using the definitions in
Equation~\ref{eq:var-multiple}, which express scenario losses relative to the
historical baseline.
\begin{equation}
\begin{aligned}
\text{VaR multiple}^{\text{(ch)}}_s &=
  \frac{\text{VaR}^{\text{(ch)}}_s}{\text{VaR}^{\text{base}}}, \\
\Delta\text{VaR}\%^{\text{(ch)}}_s &=
  100 \times
  \frac{\text{VaR}^{\text{(ch)}}_s - \text{VaR}^{\text{base}}}{|\text{VaR}^{\text{base}}|}.
\end{aligned}
\label{eq:var-multiple}
\end{equation}

An analogous definition is used for CVaR multiples.  
Unless otherwise stated, summary tables in the main text report the
\emph{linear-channel} multiples, denoted
$\text{VaR multiple}^{\text{(lin)}}_s$ and
$\text{CVaR multiple}^{\text{(lin)}}_s$, while volatility- and nonlinear-channel
results are provided alongside them in the risk tables
(e.g., Table~\ref{tab:risk-crossrun}) and the appendix.  
Table~\ref{tab:risk-crossrun} summarises the mean and standard deviation of
VaR and CVaR multiples by model, RAG, and news configuration for each channel,
while Figures~\ref{fig:cvar-by-country}, \ref{fig:inflation-vs-cvar},
\ref{fig:news-effect}, and~\ref{fig:heatmaps} visualise their distribution
across countries, channels, and RAG/news settings.  
Bootstrap confidence intervals for linear-channel multiples by
(model, RAG, news) are reported in Table~\ref{tab:bootstrap-ci}.

\subsection{Configuration Grid for Risk Evaluation}

For risk evaluation we work with the same $(\text{country}, \text{RAG}, \text{news})$ grid as for generation, yielding four configurations per country in the primary GPT-5.6-luna run.

For each configuration we compute the distribution of VaR and CVaR multiples
for Portfolios~A and~B and for each of the three stress channels, as well as
stability and cross-country consistency diagnostics:

\begin{itemize}
  \item \textbf{Country–level macro shock summary} (Table~\ref{tab:macro-summary} and Figure~\ref{fig:macro-dists}),
  \item \textbf{Portfolio A CVaR multiples by country and channel} (Figure~\ref{fig:cvar-by-country}),
  \item \textbf{News-on vs.\ news-off effects} (Figure~\ref{fig:news-effect}),
  \item \textbf{Cross–country gaps in average multiples and cross-country consistency metrics} (Table~\ref{tab:fairness}).
\end{itemize}

Cross–run comparisons of scenario severity and risk (GPT-5.6-luna vs.\ Llama-3.1-8B-Instruct) are reported in Table~\ref{tab:severity-model} and Figure~\ref{fig:model-comparison}.

\subsection{Statistical and Consistency Diagnostics}

To understand which design choices drive variation in tail risk, we run an ANOVA \cite{fisher1970statistical} on the linear-channel VaR and CVaR multiples using the following categorical factors:

\begin{itemize}
  \item \texttt{portfolio\_id} (A vs.\ B),
  \item \texttt{country} (seven G7 economies),
  \item \texttt{prompt\_variant} (30 macro narratives),
  \item \texttt{rag} (on/off),
  \item \texttt{use\_news} (on/off).
\end{itemize}

Table~\ref{tab:anova-decomp} reports, for each effect, the corresponding $p$-value and partial $\eta^2$ \cite{cohen2013statistical} for both VaR and CVaR multiples.  

In parallel, we compute a set of cross-country consistency diagnostics:

\begin{itemize}
  \item \textbf{Coverage and outliers} (Table~\ref{tab:fairness}): accepted-scenario coverage and
        full coverage, label flips under small perturbations, and outliers
        detected by both standard and robust (MAD-based) $z$-scores.

  \item \textbf{Country gaps} (Table~\ref{tab:fairness}): max--min differences
        in mean VaR and CVaR multiples across countries for each portfolio.

  \item \textbf{Linear vs.\ nonlinear channels}: the ``linear'' gaps use the
        VaR/CVaR multiples implied by the purely linear PCA mapping
        (Section~\ref{sec:methodology}), while the ``nonlinear'' gaps use
        multiples from the full nonlinear-factor channel with polynomial betas
        and $\lambda$-driven amplification. Nonlinear gaps are numerically
        small relative to linear gaps, and we therefore focus on the linear
        channel in the main text.

  \item \textbf{Flip tests} (news on vs.\ off) and outlier flags, used later in
        the results section to check that RAG/news toggles do not introduce
        unstable or systematically skewed shifts in tail risk.
\end{itemize}

These diagnostics ensure that our conclusions are not driven by a small number
of unstable configurations, and they quantify how much of the variance
in tail risk is attributable to portfolio composition, scenario design,
country, and retrieval settings.

\subsection{Compute and Reproducibility}

All experiments are run on commodity CPU instances and T4 GPU, ensuring a hardware agnostic pipeline. We log per–call latency and token usage for each model,
as well as a lightweight run manifest that hashes all critical artifacts:

\begin{itemize}
  \item WEO baselines and scenario files (CSV/JSON),
  \item cached market data (ETF prices) and derived PCA factors,
  \item FAISS indices and MiniLM weights,
  \item headline CSV snapshots for news-enabled runs,
  \item prompts and parsed JSON responses,
  \item risk tables and figure/table artifacts.
\end{itemize}

For each run we write a compact manifest
(\path{run_artifacts_index.json}) containing the run identifier,
workspace tag, model configuration, news/RAG settings, and SHA-256 hashes plus
row counts for all key files.  This ensures that any reported figure or table
can be traced back to a fixed set of documents, embeddings, scenarios, and
risk computations.

For news-enabled configurations, headline snapshots are fetched once, written
to CSV with timestamps, and reused throughout the experiment.  The manifest
records their paths and hashes, ensuring that the exact news context used at
inference time can be replayed without re-querying external APIs.

We emphasise that the pipeline is \emph{snapshot-replayable}: given the frozen
artifacts (IMF baselines, cached ETF prices, headline CSVs, MiniLM weights,
FAISS index, prompts, and global random seeds), all macro scenarios and
portfolio risk metrics can be regenerated up to floating-point and Monte Carlo
noise.  Deterministic decoding stabilizes model outputs conditional on the
retrieved context, and the deterministic historical baselines and PCA factors
eliminate external data drift, but strict hardware-agnostic bit-level
determinism is not claimed.

\section{Results}
\label{sec:results}

Unless stated otherwise, all tables and figures report the primary
GPT-5.6-luna run (\texttt{run\_A}, preset P1, results folder
\texttt{gptA\_tier\_k035\_bal\_comp\_cov50\_r0}). Results attributed to the repeat run
(\texttt{run\_B}) or to Llama-3.1-8B-Instruct (\texttt{run\_llama}) are
identified explicitly. Shocks are reported throughout as deviations in
percentage points from the April-2026 WEO projection for 2027 and from the
policy-rate baseline of Section~\ref{sec:macro-inputs}; they are never reported
as levels.

\subsection{Scenario Yield and Plausibility}
\label{sec:results-yield}

All three runs execute the full $7 \times 2 \times 2 \times 30 = 840$ cell grid.
GPT-5.6-luna returns a parseable, plausible scenario in essentially every cell
(838 accepted in each run, a rejection rate of $0.2\%$), while
Llama-3.1-8B-Instruct yields 744 accepted scenarios, a rejection rate of
$11.4\%$. Hard-gate pass rates are $0.998$, $0.998$ and $0.886$, and the worst template--configuration cell passes 6 of 7 countries for GPT
and 3 of 7 for Llama; mean soft
plausibility scores are 4.66, 4.66 and 4.51 out of 5. Without severity
bands (P4) GPT passes every cell.

The near-zero rejection rate for GPT is itself a result and not a validation of
the gate. It indicates that the hard screen---which rejects internally
contradictory macro triples and shocks failing a minimum growth shortfall---is
not binding for a model that follows the prompt closely. We therefore report the
rejection rate explicitly rather than describing the gate as a filtering
mechanism: on these runs it removes almost nothing from the proprietary model
and a non-trivial tail from the open-weight one.

\subsection{Macroeconomic Shock Distributions}
\label{sec:results-macro}

\begin{figure*}[t]
  \centering
  \includegraphics[width=\linewidth]{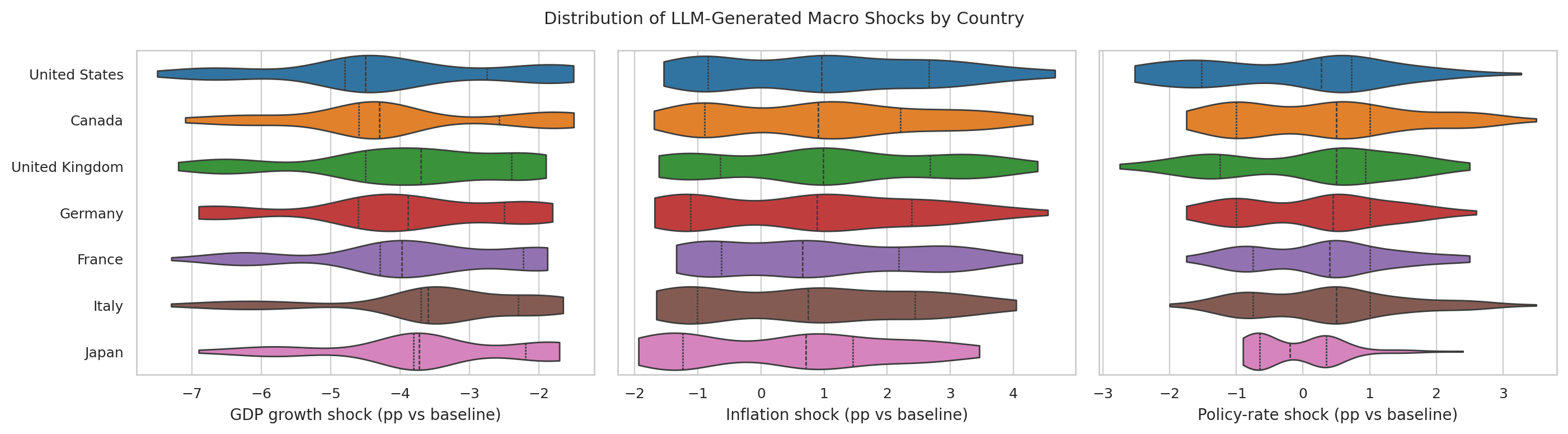}
  \caption{Violin plots of GDP, inflation and policy-rate shocks (percentage
  points, deviations from baseline) for all accepted G7 scenarios in the
  primary GPT-5.6-luna run ($N = 838$). Medians and interquartile ranges
  correspond to Table~\ref{tab:macro-summary}.}
  \label{fig:macro-dists}
\end{figure*}

Figure~\ref{fig:macro-dists} plots the full distributions and
Table~\ref{tab:macro-summary} summarises them by country for the primary
run. Growth shortfalls average $-3.85$~pp with a standard
deviation of $1.48$~pp and range from $-7.50$ to $-1.50$~pp. Inflation and
policy-rate deviations are two-sided: inflation spans $-1.94$ to $+4.66$~pp
(mean $+0.84$, s.d.\ $1.68$) and the policy rate $-2.75$ to $+3.50$~pp
(mean $+0.13$, s.d.\ $1.17$). The
two-sidedness matters. Earlier iterations of this pipeline truncated shocks at
zero before mapping them into factor space, which discarded every
disinflationary and easing scenario; the signed mapping of
Equation~\ref{eq:factor-shock} retains them.

\begin{table*}[t]
  \centering
  \small
  \caption{LLM-generated macro shocks by country, expressed as deviations in
  percentage points from the April-2026 WEO 2027 projection (GDP, inflation) and
  from the August-2026 policy rate (rate). Primary GPT-5.6-luna run,
  $N = 838$ accepted scenarios.}
  \label{tab:macro-summary}
  \begin{adjustbox}{max width=\textwidth}\begin{tabular}{lrrrrrrrrrrrrr}
    \toprule
    & \multicolumn{4}{c}{GDP shock (pp)} & \multicolumn{4}{c}{Inflation shock (pp)} & \multicolumn{4}{c}{Policy-rate shock (pp)} & \\
    \cmidrule(lr){2-5}\cmidrule(lr){6-9}\cmidrule(lr){10-13}
    Country & Mean & Std & Min & Max & Mean & Std & Min & Max & Mean & Std & Min & Max & $N$ \\
    \midrule
    Canada         & -4.03 & 1.53 & -7.10 & -1.50 & 0.93 & 1.63 & -1.70 & 4.30 &  0.28 & 1.31 & -1.75 & 3.50 & 120 \\
    France         & -3.76 & 1.43 & -7.30 & -1.88 & 0.89 & 1.57 & -1.34 & 4.14 &  0.29 & 1.03 & -1.75 & 2.50 & 119 \\
    Germany        & -4.01 & 1.50 & -6.90 & -1.80 & 0.74 & 1.72 & -1.68 & 4.54 &  0.15 & 1.10 & -1.75 & 2.60 & 120 \\
    Italy          & -3.51 & 1.39 & -7.30 & -1.65 & 0.76 & 1.70 & -1.66 & 4.04 &  0.36 & 1.11 & -2.00 & 3.50 & 120 \\
    Japan          & -3.59 & 1.35 & -6.90 & -1.70 & 0.42 & 1.64 & -1.94 & 3.46 & -0.08 & 0.66 & -0.90 & 2.40 & 120 \\
    United Kingdom & -3.92 & 1.47 & -7.20 & -1.90 & 1.09 & 1.75 & -1.62 & 4.38 &  0.11 & 1.31 & -2.75 & 2.50 & 120 \\
    United States  & -4.10 & 1.62 & -7.50 & -1.50 & 1.06 & 1.72 & -1.54 & 4.66 & -0.22 & 1.41 & -2.52 & 3.28 & 119 \\
    \bottomrule
  \end{tabular}\end{adjustbox}
\end{table*}

\subsubsection{Cross-country heterogeneity is largely baseline arithmetic}
\label{sec:results-country-artefact}

Table~\ref{tab:macro-summary} appears to show country-specific severity: the
United States and Canada receive the deepest growth shortfalls
($-4.10$ and $-4.03$~pp) while Italy and Japan receive the mildest
($-3.51$ and $-3.59$~pp). This ordering is almost entirely mechanical.

Because the shock is defined as a deviation from the WEO baseline, and because
the model selects a recession \textit{level} that varies far less across
countries than the baseline does, subtracting a larger baseline produces a
larger shock. Table~\ref{tab:country-reg} regresses the mean absolute growth
shortfall on the 2027 baseline growth rate across the seven economies.

\begin{table*}[t]
  \centering
  \small
  \caption{Regression of mean $|\Delta g|$ on the 2027 WEO baseline growth rate,
  $n = 7$ economies. Leave-one-out (LOO) columns give the range of $R^2$ and
  slope when each country is dropped in turn.}
  \label{tab:country-reg}
  \begin{adjustbox}{max width=\textwidth}\begin{tabular}{lrrrrrr}
    \toprule
    Run & Slope & SE & 95\% CI & $R^2$ & $p$ & LOO $R^2$ range \\
    \midrule
    GPT (primary) & 0.350 & 0.067 & [0.176, 0.523] & 0.843 & 0.0035 & 0.799--0.952 \\
    GPT (repeat)  & 0.342 & 0.072 & [0.158, 0.526] & 0.820 & 0.0050 & 0.761--0.944 \\
    Llama-3.1-8B  & 1.146 & 0.259 & [0.479, 1.812] & 0.796 & 0.0069 & 0.731--0.890 \\
    \midrule
    GPT (unanchored, P4) & 0.458 & 0.048 & [0.335, 0.582] & 0.948 & 0.0002 & 0.919--0.987 \\
    \bottomrule
  \end{tabular}\end{adjustbox}
\end{table*}

We report the regression fully because $n = 7$ with one predictor leaves five
residual degrees of freedom, and a $p$-value on five df carries little on its
own. The evidence rests instead on the stability of the fit: dropping any single
country leaves $R^2$ between $0.731$ and $0.952$ and the slope within
$\pm 21\%$ of its full-sample value in every run. The United States has the
highest leverage ($h_{ii} = 0.50$), and excluding it still yields
$R^2 = 0.820$. The implied severity \textit{levels} span only $1.01$~pp across
the G7 (from $-2.00\%$ for the United States to $-3.01\%$ for Italy), whereas
the baselines themselves span $1.60$~pp.

We therefore do not interpret the country effect in
Table~\ref{tab:anova-decomp} ($\eta^2 \approx 0.09$--$0.14$ within Portfolio~A) as evidence that
the models perform country-specific vulnerability assessment. The correct
reading is the opposite: conditional on a prescribed narrative, these models
choose a broadly similar recession depth irrespective of the economy in
question, and the heterogeneity visible in delta space is inherited from the IMF
baseline. Any stress-testing pipeline that reports shocks as deviations from a
heterogeneous baseline is exposed to this artefact, and we recommend that the
baseline regression be reported alongside any claimed country effect. Without
the severity bands (P4) the baseline explains even more of the country pattern
($R^2 = 0.948$).

Ablation S8 provides a direct test that does not rely on seven data points. It
re-runs the Llama risk layer on the model's own \emph{stated} changes instead of
level-minus-baseline deviations, which removes the baseline subtraction from
the shock altogether. The country effect on Portfolio~A linear-channel VaR
multiples then falls from $\eta^2 = 0.087$ ($p < 0.001$) to $0.013$
($p = 0.13$), while the prompt effect is almost unchanged ($0.398$ against
$0.361$). Once the arithmetic of the baseline is taken out, the open-weight
model's scenarios carry no detectable country signal. The same test is
uninformative for GPT, whose stated changes coincide with its computed
deviations in every scenario.

\subsection{Severity Tiers and Prompt Anchoring}
\label{sec:results-tiers}

Each prompt template is written for one severity tier (contained, severe,
systemic) and supplies that tier's indicative, overlapping magnitude band. The
tier itself is not given as a field to copy: the model must return the
\texttt{severity\_tier} its rationale describes.
Table~\ref{tab:tiers} reports the realised shocks by declared tier.

\begin{table*}[t]
  \centering
  \small
  \caption{Realised macro shocks by self-declared severity tier. Values are
  means with standard deviations in parentheses. Tier consistency is the share
  of scenarios whose declared tier equals the modal tier of their template. One Llama scenario declares no
  tier.}
  \label{tab:tiers}
  \begin{adjustbox}{max width=\textwidth}\begin{tabular}{llrrrr}
    \toprule
    Run & Tier & $N$ & GDP (pp) & Inflation (pp) & Rate (pp) \\
    \midrule
    \multirow{3}{*}{GPT-5.6-luna}
      & contained & 224 & -1.98 (0.25) &  0.72 (1.14) &  0.05 (0.65) \\
      & severe    & 472 & -3.99 (0.49) &  1.20 (1.84) &  0.41 (1.21) \\
      & systemic  & 142 & -6.32 (0.58) & -0.14 (1.46) & -0.70 (1.25) \\
    \midrule
    \multirow{3}{*}{Llama-3.1-8B}
      & contained & 236 & -2.84 (1.76) &  0.78 (1.38) &  0.75 (1.11) \\
      & severe    & 388 & -3.96 (1.99) &  1.07 (1.70) &  0.80 (1.47) \\
      & systemic  & 119 & -5.43 (2.91) &  0.63 (2.13) &  0.99 (2.25) \\
    \midrule
    \multicolumn{6}{l}{\textit{Tier consistency:} GPT 0.996 (29 of 30 templates fully consistent), Llama 0.888 (5 of 30)} \\
    \bottomrule
  \end{tabular}\end{adjustbox}
\end{table*}

Declared severity orders the realised growth shock monotonically in both models,
which is the coherence property the design intends. However, two features
prevent this from being read as validation of the models' judgement.

First, the labels are near-deterministic functions of the prompt. GPT assigns every scenario from a
template the same tier in $99.6\%$ of cases (29 of 30 templates without a single
exception), and both GPT draws assign the same tier to every template, so
\texttt{severity\_tier} is close to a relabelling of \texttt{prompt\_variant}
and carries almost no independent information. Llama is less consistent
($88.8\%$, as low as $52\%$ for a single template, and only 5 templates without
exception), which makes its tier assignment marginally informative; its modal
tier nevertheless agrees with GPT's on 29 of the 30 templates. GPT's modal tiers
reproduce the intended allocation of 8 contained, 17 severe and 5 systemic
templates exactly.

Second, and more consequentially, GPT's within-tier standard deviations
($0.25$, $0.49$, $0.58$~pp) are small relative to the width of the indicative
bands ($2$, $4$ and $5$~pp respectively), and the tier means lie within $0.2$~pp
of each band midpoint despite an explicit instruction not to default to the
middle of the range. GPT is therefore anchoring on the prompted magnitude rather
than deriving it from the mechanism it describes. Llama does not: its within-tier
dispersion is $1.76$--$2.91$~pp, four to seven times larger.

\begin{figure*}[t]
  \centering
  \includegraphics[width=0.85\linewidth]{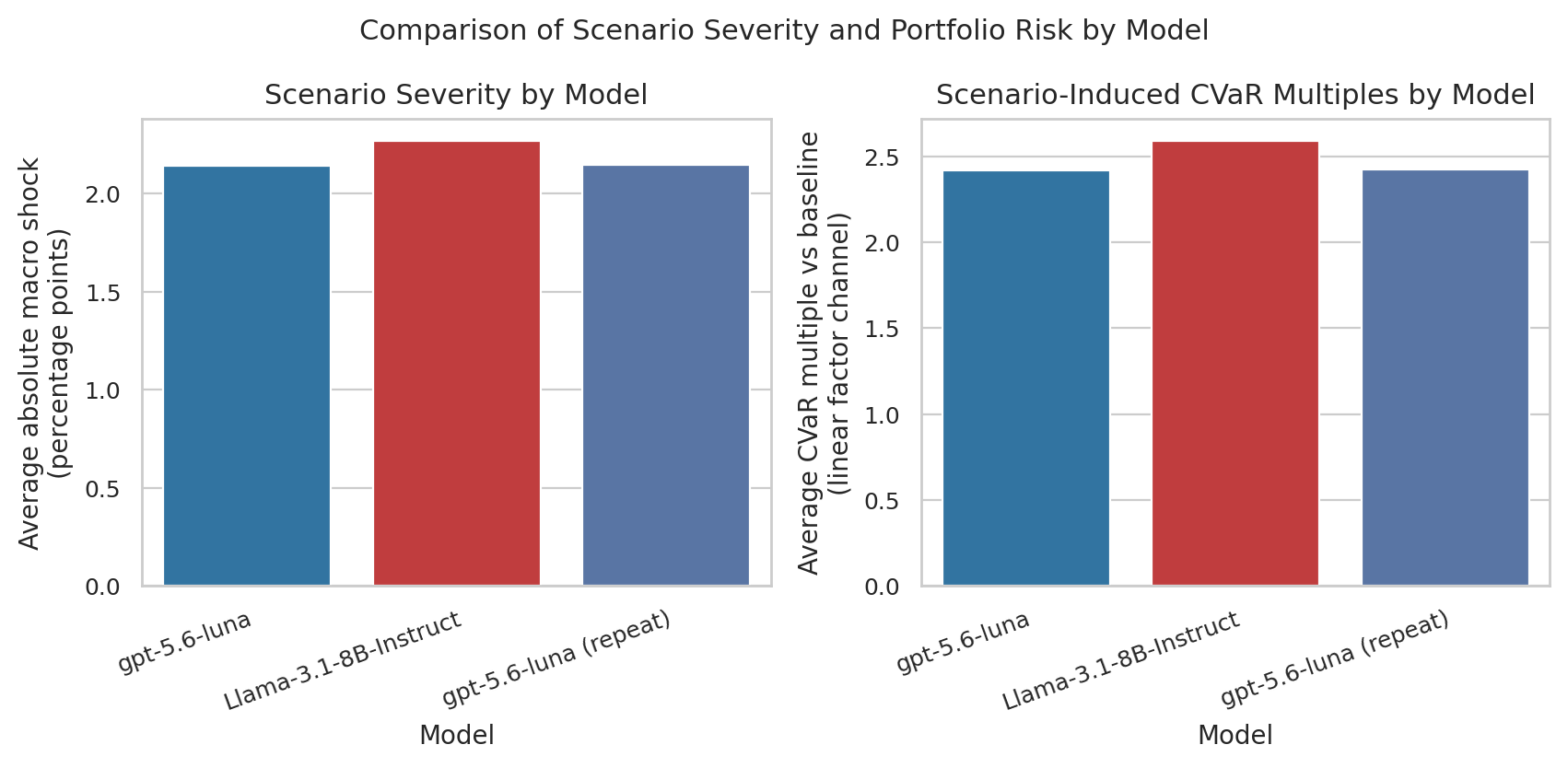}
  \caption{Average absolute macro shock severity (left) and linear-channel CVaR
  multiples (right) for the primary GPT, repeat GPT and Llama runs, pooling over
  countries and configurations.}
  \label{fig:model-comparison}
\end{figure*}

Table~\ref{tab:severity-model} summarises unconditional severity by run and
Figure~\ref{fig:model-comparison} contrasts the two model families on average
shock magnitude and induced CVaR multiples.

This is the central caveat of the paper and we state it plainly. In the
configuration studied here, the numerical severity of a GPT-generated scenario
is predominantly a property of the prompt, not of the model's macroeconomic
reasoning. That is not automatically a defect---supervisory frameworks such as
CCAR and the EBA exercises likewise \textit{prescribe} severity---but it means
the model's contribution should be located in the narrative, the transmission
mechanism and the sector attribution, not in the calibration of magnitude.
Section~\ref{sec:results-ablations} shows directly what happens when the bands
are removed.

\begin{table*}[t]
  \centering
  \small
  \caption{Scenario severity by run: unconditional macro deviation magnitudes
    across all countries and configurations.}
  \label{tab:severity-model}
  \begin{adjustbox}{max width=\textwidth}\begin{tabular}{lrrrrr}
    \toprule
    Run & $\Delta g$ & $\Delta \pi$ & $\Delta r$ & Mean $|\Delta \mathbf{M}|$ & $N$ \\
    \midrule
    GPT-5.6-luna (primary)   & -3.85 & 0.84 & 0.13 & 2.14 & 838 \\
    GPT-5.6-luna (repeat) & -3.85 & 0.85 & 0.12 & 2.15 & 838 \\
    Llama-3.1-8B-Instruct & -3.84 & 0.91 & 0.82 & 2.27 & 744 \\
    \bottomrule
  \end{tabular}\end{adjustbox}
\end{table*}

\subsection{Are the Numbers Coupled to the Narrative?}
\label{sec:results-grounding}

The pipeline assumes that a scenario's macro triple and its free-text rationale
describe the same event. We test this directly. The DeBERTa regime classifier
reads only the rationale and returns a continuous severity score; correlating
that score with the realised growth shock measures whether narrative severity
and numeric severity move together.

Pooled across all accepted scenarios the correlation is weak and correctly
signed for GPT ($r = -0.117$ and $-0.106$ across the two runs) and close to
zero for Llama ($r = -0.055$). For Llama the model's \emph{stated} change tracks
the narrative somewhat better than the level-minus-baseline deviation
($r = -0.120$), which is the diagnostic that motivated ablation S8. Removing the
severity bands does not restore the coupling: in the unanchored run the pooled
correlation is $+0.014$. Because the tier design induces a common
component in both quantities, the sharper test is within tier. That test,
however, conditions on a variable whose within-tier dispersion is a fraction of
its unconditional dispersion, and a near-zero conditional correlation is exactly
what \emph{range restriction} predicts even under a strong underlying
association. We therefore report both the observed and the restriction-corrected
correlation in Table~\ref{tab:grounding}, using the Thorndike Case~II correction
for direct restriction on the outcome variable.

\begin{table*}[t]
  \centering
  \small
  \caption{Association between the realised growth deviation and the narrative
  regime score, within severity tier. ``Ratio'' is the unconditional standard
  deviation of $\Delta g$ divided by its within-tier standard deviation; a ratio
  near one means the tier imposes little restriction. $r_{\text{corr}}$ applies
  the Thorndike Case~II correction.}
  \label{tab:grounding}
  \begin{adjustbox}{max width=\textwidth}\begin{tabular}{llrrrrr}
    \toprule
    Run & Tier & $N$ & SD($\Delta g$) & Ratio & $r_{\text{obs}}$ & $r_{\text{corr}}$ \\
    \midrule
    \multirow{3}{*}{GPT (primary)}
      & contained & 224 & 0.255 & 5.82 & $-0.264$ & $-0.847$ \\
      & severe    & 472 & 0.489 & 3.03 & $+0.087$ & $+0.256$ \\
      & systemic  & 142 & 0.581 & 2.55 & $-0.051$ & $-0.129$ \\
    \midrule
    \multirow{3}{*}{GPT (repeat)}
      & contained & 224 & 0.260 & 5.72 & $-0.148$ & $-0.652$ \\
      & severe    & 472 & 0.465 & 3.20 & $+0.067$ & $+0.209$ \\
      & systemic  & 142 & 0.586 & 2.54 & $-0.066$ & $-0.166$ \\
    \midrule
    \multirow{3}{*}{Llama-3.1-8B}
      & contained & 236 & 1.755 & 1.29 & $+0.028$ & $+0.036$ \\
      & severe    & 388 & 1.991 & 1.14 & $-0.033$ & $-0.038$ \\
      & systemic  & 119 & 2.914 & 0.78 & $+0.082$ & $+0.064$ \\
    \bottomrule
  \end{tabular}\end{adjustbox}
\end{table*}

The correction changes the reading materially in one place and confirms it
elsewhere. For the \emph{contained} tier under GPT the restriction ratio is
$5.8$, and correcting for it turns an apparently modest $r = -0.26$ into
$r_{\text{corr}} = -0.85$. Within that tier, narrative and numeric severity are
in fact strongly coupled, and our earlier reading of this cell was wrong.

For the severe and systemic tiers---which together account for $73\%$ of
accepted scenarios---the restriction ratios are $2.5$--$3.2$; the corrected correlations are small
and of mixed sign ($+0.256$, $-0.129$, $+0.209$, $-0.166$), and none of the
underlying observed correlations is significant at the $5\%$ level. In the
repeat run the contained-tier correlation is weaker ($r = -0.15$,
$r_{\text{corr}} = -0.65$), so the one strong association is itself not stable
across draws. The Llama run provides the cleaner test: its within-tier
dispersion is close to its unconditional dispersion (ratios $0.78$--$1.29$), so
restriction is negligible and no correction is needed, and its correlations are
near zero in every tier. The decoupling finding therefore survives where the
data can support it and fails in the one tier where restriction was severe.

Because correlations on a compressed instrument are hard to interpret, we also
report slopes on the natural scale. Over the observed regime-score range of
roughly $0.45$ units, the pooled slope for the primary run is $-1.76$~pp per
unit ($95\%$ CI $[-2.78, -0.74]$), i.e.\ about $0.8$~pp of growth deviation
across the entire range of narrative severity the classifier can express. Within
the severe tier the slope is $+0.42$~pp per unit ($95\%$ CI $[-0.02, +0.86]$)
and within the systemic tier $-0.30$~pp ($95\%$ CI $[-1.28, +0.68]$): both
intervals comfortably contain zero and exclude any economically meaningful
association.

Two caveats bound this analysis. The classifier never returns \texttt{normal}
for any scenario---the maximum $p_{\text{normal}}$ observed is $0.09$ for GPT
and $0.42$ for Llama---so the effective score range is $[0.45, 0.99]$ rather
than $[0,1]$. And the classifier is a general-purpose NLI model with no
validation against expert judgements of macro-financial severity; dispersion is
not validity, and attenuation from measurement error biases correlations toward
zero. The correct statement of this finding is therefore narrower than
``narrative and numeric severity are decoupled'': it is that
\emph{DeBERTa-derived narrative severity scores are weakly associated with
numeric severity in the severe and systemic tiers, and for the open-weight model
throughout}. Establishing the stronger claim requires human severity ratings on
a stratified sample with reported inter-rater agreement, which we identify as
required future work in Section~\ref{sec:disc-future} rather than claiming here.

The finding has a direct architectural consequence. The regime severity index
$\lambda_s$ that mixes the calm and stress-period covariance matrices is derived from
the rationale, while the factor drift is derived from the macro triple. If the
two are near-orthogonal, the volatility channel and the factor channels are
driven by substantially independent signals rather than by a single coherent
notion of scenario severity. Practitioners building comparable pipelines should
either enforce coupling explicitly or report the two channels as measuring
different things.

\subsection{Portfolio Tail-Risk Multiples}
\label{sec:results-risk}

Table~\ref{tab:risk-crossrun} reports VaR and CVaR multiples relative to the
63-day historical bootstrap baseline, by model, retrieval and news setting, for
all three stress channels at the production calibration
($3.5\%$ equity decline per percentage point of growth shortfall,
Section~\ref{sec:calibration}). Bootstrap confidence intervals for the linear
channel appear in Table~\ref{tab:bootstrap-ci}.

\begin{table*}[t]
  \centering
  \small
  \caption{Portfolio tail risk by retrieval (RAG) and news configuration for the
  primary GPT-5.6-luna run, pooled across countries and both portfolios. Values are VaR/CVaR multiples
  relative to the historical bootstrap baseline for each stress channel, at the
  production calibration. Pooled over all configurations, the current runs give
  linear-channel multiples of $4.06/2.44$ (GPT primary), $4.06/2.44$ (GPT repeat)
  and $4.28/2.56$ (Llama) for VaR/CVaR, and volatility-channel multiples of
  $2.82/1.82$, $2.82/1.83$ and $2.92/1.89$. Per-configuration linear-channel
  intervals for all three runs are in Table~\ref{tab:bootstrap-ci}.}
  \label{tab:risk-crossrun}
  \resizebox{\textwidth}{!}{%
  \begin{adjustbox}{max width=\textwidth}\begin{tabular}{lllrrrrrrrrrrrrr}
    \toprule
    & & & \multicolumn{4}{c}{Volatility channel} & \multicolumn{4}{c}{Linear channel} & \multicolumn{4}{c}{Nonlinear channel} & \\
    \cmidrule(lr){4-7}\cmidrule(lr){8-11}\cmidrule(lr){12-15}
    & & & \multicolumn{2}{c}{VaR} & \multicolumn{2}{c}{CVaR} & \multicolumn{2}{c}{VaR} & \multicolumn{2}{c}{CVaR} & \multicolumn{2}{c}{VaR} & \multicolumn{2}{c}{CVaR} & \\
    Model & RAG & News & Mean & Std & Mean & Std & Mean & Std & Mean & Std & Mean & Std & Mean & Std & $N$ \\
    \midrule
    GPT-5.6-luna & Off & Off & 2.80 & 0.30 & 1.82 & 0.18 & 4.05 & 0.67 & 2.43 & 0.35 & 4.05 & 0.68 & 2.44 & 0.36 & 420 \\
    GPT-5.6-luna & Off & On  & 2.83 & 0.30 & 1.83 & 0.18 & 4.07 & 0.68 & 2.45 & 0.36 & 4.07 & 0.68 & 2.45 & 0.36 & 420 \\
    GPT-5.6-luna & On  & Off & 2.82 & 0.31 & 1.83 & 0.18 & 4.07 & 0.69 & 2.44 & 0.36 & 4.07 & 0.68 & 2.45 & 0.36 & 418 \\
    GPT-5.6-luna & On  & On  & 2.81 & 0.29 & 1.82 & 0.17 & 4.05 & 0.68 & 2.43 & 0.36 & 4.05 & 0.67 & 2.43 & 0.35 & 418 \\
    \bottomrule
  \end{tabular}\end{adjustbox}%
  }
\end{table*}

Three observations follow.

\textbf{Retrieval and news cannot move the result, by construction.} We state
the architectural argument first because it determines how the statistics should
be read. Text reaches the risk translator only through the curvature
term of the nonlinear channel, and that term contributes less than $0.6\%$ of the VaR or CVaR estimate. Everything else that retrieval and news could change must pass through
the macro numbers the model emits. A
signal routed exclusively through a channel of that magnitude cannot move the
output whatever its content. The retrieval and news rows in
Table~\ref{tab:anova-decomp} and Table~\ref{tab:cluster-robust} should therefore
be read as \emph{confirming a structural constraint we imposed}, not as testing
a live hypothesis about whether grounding matters.

\begin{figure*}[t]
  \centering
  \includegraphics[width=0.66\linewidth]{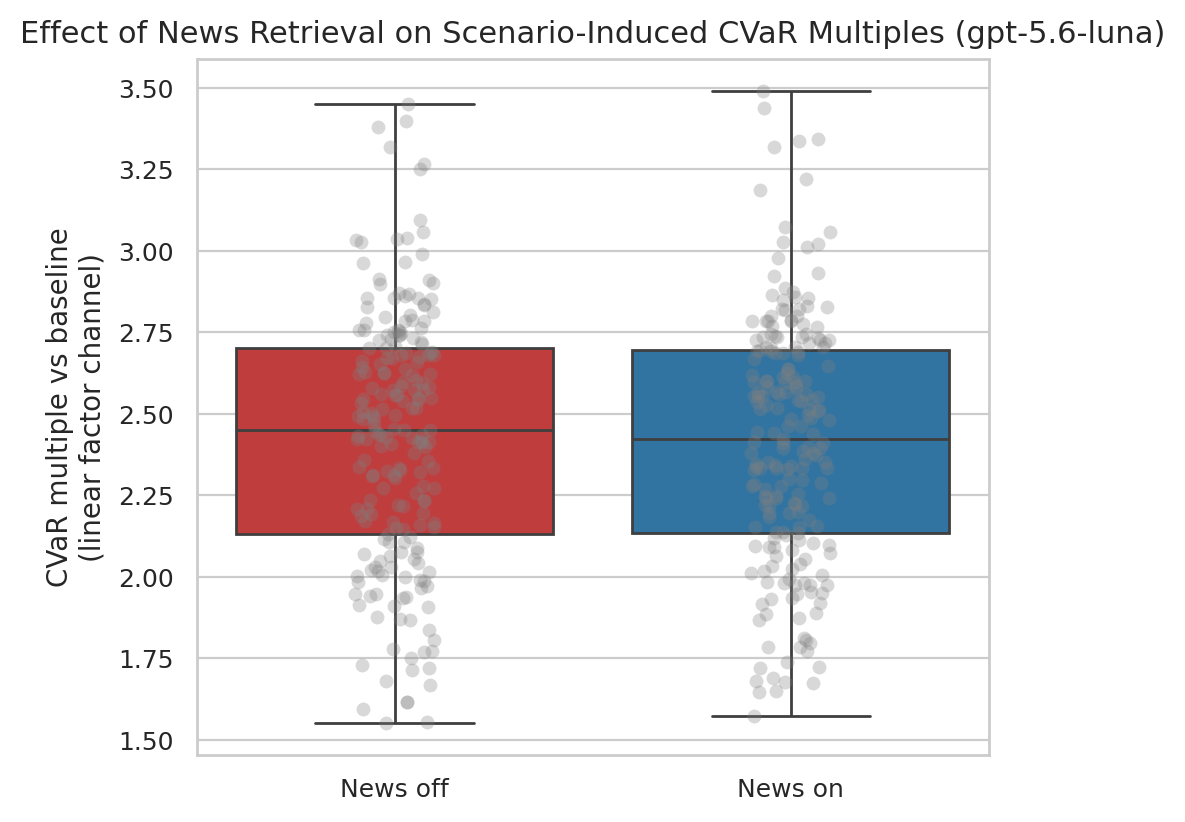}
  \caption{Distribution of linear-channel CVaR multiples with news retrieval on
  versus off (RAG enabled, primary run). The distributions are visually
  indistinguishable, consistent with the null ANOVA effect in
  Table~\ref{tab:anova-decomp}.}
  \label{fig:news-effect}
\end{figure*}

With that framing, the numbers are as expected for GPT under tiered bands:
retrieval and news are null on Portfolio~A in both runs ($p \geq 0.10$,
$\eta^2 \leq 0.003$; Table~\ref{tab:anova-decomp}), across the four
configurations the linear VaR multiple varies from $3.920$ to $3.952$---a spread
of $0.8\%$ against a within-cell standard deviation of $0.71$---and the bootstrap
intervals overlap completely. The emitted-number route is, however, not always
closed. News has a small but statistically significant effect for Llama
($\eta^2 = 0.008$, $p = 0.02$ on Portfolio~A) and for GPT once the severity bands
are removed ($\eta^2 = 0.012$, $p = 0.002$; Section~\ref{sec:results-ablations}).
Grounding reaches the risk layer exactly where the prompt leaves the magnitude
free to move; retrieval, whose context differs across countries only in a few
WEO numbers, is null in every run. This is emphatically \emph{not} evidence
that retrieval is ineffective for scenario generation in general. It is evidence
that our architecture gives retrieval almost no opportunity to affect the
quantity we measure, which is a design limitation we report rather than a
finding about RAG. Figure~\ref{fig:news-effect} shows the two news
distributions superimposed.

\textbf{The nonlinear channel is empty.} Correctly specified as an increment
over the linear drift rather than a rival estimate of it, the curvature
contribution is $+0.02\%$ and $+0.01\%$ on VaR and CVaR for GPT, and
$-0.53\%$ and $-0.46\%$---that is, slightly risk-\textit{reducing}---for Llama.
Polynomial curvature in daily PCA factors carries essentially no incremental
tail risk at this horizon. The sign is the opposite of what an amplification
channel is intended to deliver.

\textbf{Llama produces more severe and more dispersed risk, subject to a
selection caveat.} Pooled over both portfolios, its linear VaR multiple averages $4.28$ against
GPT's $4.06$ ($+5\%$) and its CVaR multiple $2.56$ against $2.44$, with a
Portfolio~A standard deviation of $1.03$ against $0.70$. The two models emit the
same mean growth shortfall ($-3.84$ against $-3.85$~pp;
Table~\ref{tab:severity-model}); the gap comes from Llama's much larger mean
rate deviation ($+0.82$ against $+0.13$~pp) and its wider dispersion. Part of the gap is an
artefact of Llama's inconsistent arithmetic: when its own stated changes are used
as shocks (S8) the pooled multiples fall to $3.99$ and $2.41$, marginally below
GPT's, although S8 also re-gates the corpus (780 accepted) and so is not an
identical scenario set.
This comparison is made on accepted scenarios only, and the two models are
filtered at very different rates ($744/840$ versus $838/840$), so part of the
difference may reflect gate-induced selection rather than model capability. Of Llama's 96
rejections, 94 fail the plausibility gate and 2 carry invalid stated changes. The
deviation floor binds for Llama---its least severe accepted scenario sits at
$\Delta g = -0.503$~pp, just past the threshold---which means the gate is
truncating the mild end of its distribution while leaving GPT's untouched. A
full pre- versus post-filter comparison with rejection reasons by category would
be required to separate the two, and we flag it as a limitation rather than
claiming a clean capability difference. This follows directly from its wider
scenario distribution (Section~\ref{sec:results-tiers}) rather than from any
difference in the risk translator, which is identical across runs.

\paragraph{The machinery shifts the mean, not the tail.} For a paper about tail
risk this deserves direct statement. The historical baseline has
$\text{CVaR}/\text{VaR} = 0.0911/0.0490 = 1.86$; the scenario distributions have
$1.15$ in the linear channel and $1.23$ in the volatility channel, with standard
deviations under $0.04$. The simulated scenarios are markedly \emph{less}
fat-tailed than the empirical baseline, because a Gaussian covariance simulation
with a shifted drift reproduces a near-Gaussian tail shape whatever the drift.
What the pipeline produces is a severe central outcome, not a heavy tail in the
sense the stress-testing literature intends. The reported multiples should be
read as ``how far the distribution moves'' and not as ``how much fatter the tail
becomes''.

\paragraph{The severity index and the volatility channel.} The realised
$\lambda_s$ (Equation~\ref{eq:lambda}) spans $0.293$ to $0.718$ for the primary
run with a median of $0.496$, and $51.6\%$ of scenarios fall below $0.5$; for
Llama the range is $0.238$--$0.972$ with $38.0\%$ below $0.5$. The covariance
mixture is therefore not pinned to the stress regime, but neither does it ever
approach the calm regime: no scenario receives a mixture weight below $0.23$.
This is a direct consequence of the classifier never emitting \texttt{normal}
(Section~\ref{sec:results-grounding}), and it explains why volatility-channel
multiples are tightly clustered ($2.72 \pm 0.28$ for the primary run on
Portfolio~A, range $2.16$--$3.62$). The volatility channel is reporting a largely fixed consequence
of the $\lambda$ floor rather than a scenario-specific quantity, and should be
read that way.

\paragraph{Sources of uncertainty.} The bootstrap intervals in
Table~\ref{tab:bootstrap-ci} resample scenario-level multiples and therefore
capture scenario-generation uncertainty only. Two further sources are present.
Monte Carlo error in each 20{,}000-path VaR estimate contributes a relative
standard error of at most about $0.9\%$ of the estimate under a Gaussian
approximation to the simulated loss distribution, roughly an order of
magnitude below the scenario dispersion of $17.9\%$ of the mean (primary run) or
$24.4\%$ (Llama). Parameter uncertainty in the PCA loadings and covariance
matrices is not quantified anywhere in this study and we do not claim it is
negligible.

\begin{figure*}[t]
  \centering
  \includegraphics[width=\linewidth]{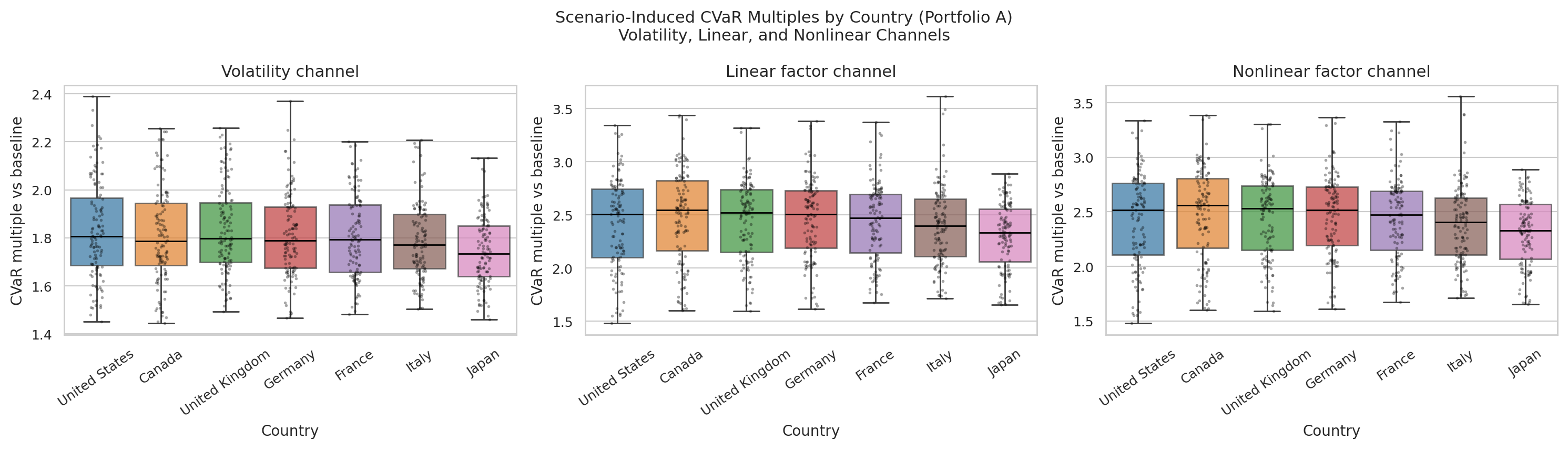}
  \caption{Linear-channel CVaR multiples for Portfolio~A by country, pooling
  over retrieval and news configurations in the primary run. Values are
  multiples relative to the historical-bootstrap baseline.}
  \label{fig:cvar-by-country}
\end{figure*}

\begin{figure*}[t]
  \centering
  \includegraphics[width=0.72\linewidth]{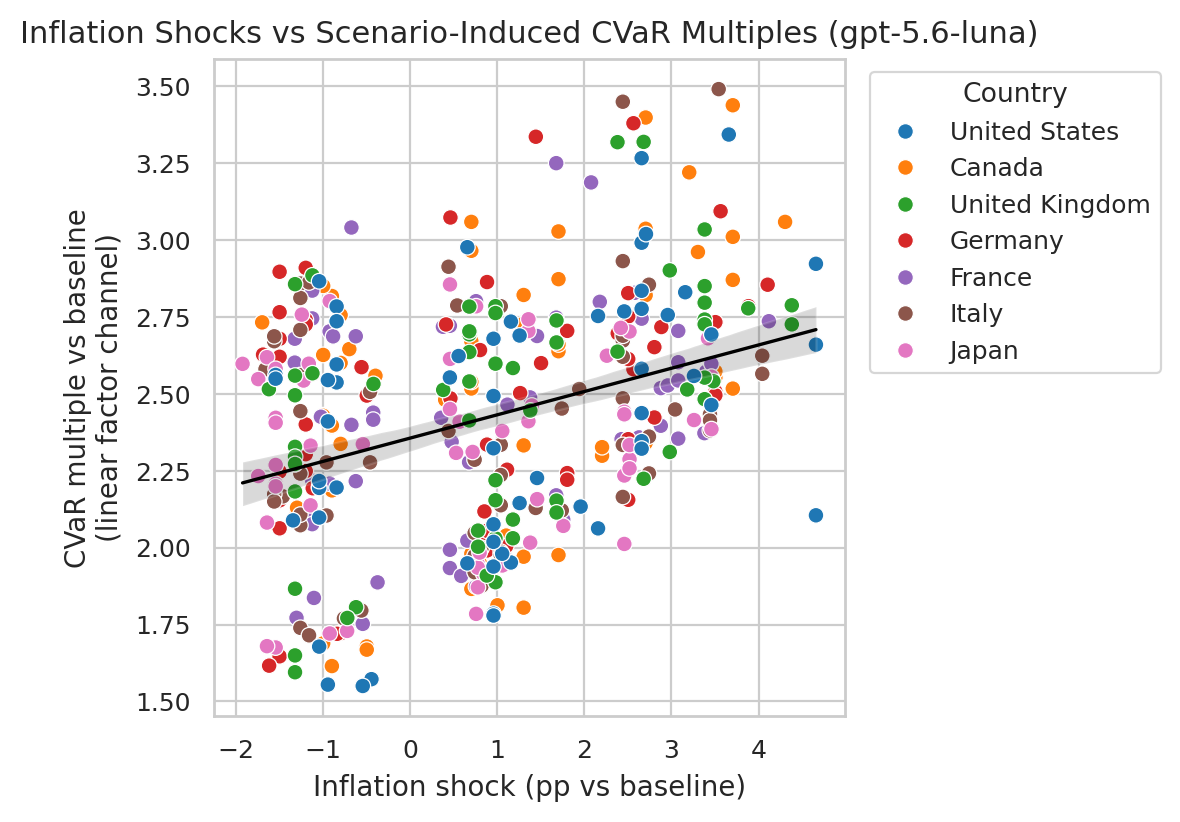}
  \caption{Scenario-induced CVaR multiples for Portfolio~A plotted against
  inflation deviations across the three translation channels. The weak
  association reflects that tail risk depends on the joint shock vector, regime
  mixing and portfolio composition rather than on inflation alone.}
  \label{fig:inflation-vs-cvar}
\end{figure*}

Figure~\ref{fig:cvar-by-country} disaggregates the CVaR multiple by country and
channel, and Figure~\ref{fig:inflation-vs-cvar} plots it against the inflation
deviation. The association with inflation alone is weak, which is expected: tail
risk depends on the joint shock vector, the regime mixture and portfolio
composition rather than on any single macro variable.

\paragraph{Within-portfolio results.} Because we argue in
Section~\ref{sec:results-anova} that pooling the two portfolios violates
independence, Table~\ref{tab:within-portfolio} reports the linear-channel
multiples separately by portfolio. On Portfolio~A the four retrieval and news cells differ by
at most $0.03$ multiples for GPT and $0.18$ for Llama, against within-cell
standard deviations of $0.71$ and $1.03$.

\begin{table*}[t]
  \centering
  \small
  \caption{Linear-channel multiples reported within portfolio rather than
  pooled. Standard deviations in parentheses.}
  \label{tab:within-portfolio}
  \begin{adjustbox}{max width=\textwidth}\begin{tabular}{lllrrr}
    \toprule
    Run & Portfolio & Config & VaR mult. & CVaR mult. & $N$ \\
    \midrule
    \multirow{4}{*}{GPT (primary)}
      & A & all             & 3.936 (0.704) & 2.420 (0.384) & 838 \\
      & A & RAG on, news on & 3.927 (0.706) & 2.415 (0.383) & 209 \\
      & B & all             & 4.179 (0.631) & 2.456 (0.331) & 838 \\
      & B & RAG on, news on & 4.165 (0.625) & 2.448 (0.327) & 209 \\
    \midrule
    \multirow{4}{*}{GPT (repeat)}
      & A & all             & 3.941 (0.700) & 2.423 (0.381) & 838 \\
      & A & RAG on, news on & 3.957 (0.711) & 2.433 (0.386) & 210 \\
      & B & all             & 4.186 (0.627) & 2.460 (0.328) & 838 \\
      & B & RAG on, news on & 4.194 (0.627) & 2.464 (0.327) & 210 \\
    \midrule
    \multirow{4}{*}{Llama-3.1-8B}
      & A & all             & 4.243 (1.033) & 2.590 (0.558) & 744 \\
      & A & RAG on, news on & 4.240 (1.107) & 2.587 (0.598) & 184 \\
      & B & all             & 4.311 (0.928) & 2.530 (0.481) & 744 \\
      & B & RAG on, news on & 4.248 (0.968) & 2.497 (0.503) & 184 \\
    \bottomrule
  \end{tabular}\end{adjustbox}
\end{table*}

\begin{table*}[t]
  \centering
  \small
  \caption{Mean linear- and volatility-channel multiples for Portfolio~A against
  each of the three benchmarks in Table~\ref{tab:baseline-var}. The denominator
  changes the headline linear-channel VaR figure by more than $60\%$.}
  \label{tab:multi-baseline}
  \begin{adjustbox}{max width=\textwidth}\begin{tabular}{llrrrr}
    \toprule
    & & \multicolumn{2}{c}{Linear channel} & \multicolumn{2}{c}{Volatility channel} \\
    \cmidrule(lr){3-4}\cmidrule(lr){5-6}
    Run & Benchmark & VaR & CVaR & VaR & CVaR \\
    \midrule
    \multirow{3}{*}{GPT (primary)}
      & Bootstrap  & 3.94 & 2.42 & 2.72 & 1.80 \\
      & EWMA       & 2.55 & 2.32 & 1.76 & 1.73 \\
      & GARCH--$t$ & 2.43 & 2.02 & 1.68 & 1.51 \\
    \midrule
    \multirow{3}{*}{Llama-3.1-8B}
      & Bootstrap  & 4.24 & 2.59 & 2.81 & 1.87 \\
      & EWMA       & 2.74 & 2.48 & 1.82 & 1.79 \\
      & GARCH--$t$ & 2.62 & 2.16 & 1.74 & 1.56 \\
    \bottomrule
  \end{tabular}\end{adjustbox}
\end{table*}

\subsection{Calibration Sensitivity}
\label{sec:results-sensitivity}

\begin{table*}[t]
  \centering
  \small
  \caption{Calibration sensitivity. $\kappa$ is the assumed equity decline per
  percentage point of growth shortfall over the 63-day horizon. Volatility-channel
  multiples are invariant to $\kappa$ by construction. Multiples are pooled over
  both portfolios. The GPT-primary rows come from the separate ablation runs
  S1, P1 and S2; the $\kappa = 3.5\%$ rows of the other runs are exact, and
  their $2\%$ and $5\%$ rows are derived from the in-run calibration sweep
  (accurate to about $\pm 0.01$).}
  \label{tab:sensitivity}
  \begin{adjustbox}{max width=\textwidth}\begin{tabular}{llrrrr}
    \toprule
    Run & $\kappa$ & VaR (vol.) & CVaR (vol.) & VaR (lin.) & CVaR (lin.) \\
    \midrule
    \multirow{3}{*}{GPT (primary)}
      & 2.0\% & 2.816 & 1.824 & 3.481 & 2.144 \\
      & 3.5\% & 2.816 & 1.824 & 4.058 & 2.438 \\
      & 5.0\% & 2.816 & 1.824 & 4.605 & 2.716 \\
    \midrule
    \multirow{3}{*}{GPT (repeat)}
      & 2.0\% & 2.823 & 1.828 & 3.491 & 2.149 \\
      & 3.5\% & 2.823 & 1.828 & 4.064 & 2.442 \\
      & 5.0\% & 2.823 & 1.828 & 4.608 & 2.719 \\
    \midrule
    \multirow{3}{*}{Llama-3.1-8B}
      & 2.0\% & 2.916 & 1.886 & 3.660 & 2.247 \\
      & 3.5\% & 2.916 & 1.886 & 4.277 & 2.560 \\
      & 5.0\% & 2.916 & 1.886 & 4.853 & 2.853 \\
    \bottomrule
  \end{tabular}\end{adjustbox}
\end{table*}

The level of every reported multiple depends on
$\kappa$, the assumed equity response to a one percentage point growth
shortfall. Reporting a single value would present a calibration choice as an
empirical finding, so we re-run the risk layer across
$\kappa \in \{2\%, 3.5\%, 5\%\}$ (Table~\ref{tab:sensitivity}).

The response is close to exactly linear in $\kappa$ ($R^2 = 0.9998$ for the
primary run). We note explicitly that this linearity is \emph{not} evidence of model
validity: the translator is linear in $\kappa$ by construction, so near-perfect
linearity is arithmetic rather than an empirical finding. It does confirm the
narrow point that no clipping or threshold artefact intrudes over this range,
and it licenses reporting an interval rather than three disconnected point
estimates. We therefore report
LLM-induced linear-channel severity as
$3.5$--$4.6\times$ VaR and $2.1$--$2.7\times$ CVaR for GPT, and
$3.7$--$4.9\times$ and $2.2$--$2.9\times$ for Llama. The empirical-$\kappa$
ablation (S3) is not a row of Table~\ref{tab:sensitivity}: its estimate is
unusable (Section~\ref{sec:calibration}) and the run falls back to the central
value.

Section~\ref{sec:crisis-envelopes} places these against historical crisis
envelopes. At the central calibration the multiples sit close to the Global
Financial Crisis envelope measured against the same unconditional baseline. We
stress that this proximity is a consequence of the calibration choice and not an
independent finding: at $\kappa = 2\%$ the scenarios are materially milder than
the GFC, and at $\kappa = 5\%$ materially more severe.

\subsection{Reproducibility Across Repeat Runs}
\label{sec:results-reproducibility}

The two GPT runs use identical prompts, identical retrieval artefacts, an
identical decode configuration and the same frozen news snapshot. They are
repeat draws, not a temperature ablation: the GPT-5 family accepts only
$\text{temperature} = 1$, so strict greedy decoding is unavailable and any
claim of deterministic generation would be false.

At the level of individual scenarios the two runs differ substantially. Matching
on the 836 (country, prompt, RAG, news) cells accepted in both runs, only
$53.9\%$ of growth shocks, $52.9\%$ of inflation shocks and $57.8\%$ of
policy-rate shocks are bit-identical, and only $57.2\%$ of scenario titles match
exactly. Mean absolute differences are $0.16$--$0.17$~pp with maxima of $1.30$,
$2.60$ and $3.80$~pp respectively.

At the level of reported results the two runs are almost indistinguishable.
Both runs accept 838 scenarios; their mean inflation deviations are $0.842$
and $0.845$~pp, their mean rate deviations $0.126$ and $0.117$~pp, and their mean
emitted growth rates $-2.635\%$ and $-2.642\%$. Cross-run correlations are
$0.979$, $0.975$ and $0.937$ for the three shocks; declared severity tiers
agree in $99.5\%$ of cells; and the pooled mean multiples differ by at most
$0.007\times$ in any channel (linear VaR $4.058$ versus $4.064$, linear CVaR
$2.438$ versus $2.442$, volatility VaR $2.816$ versus $2.823$), well inside the
bootstrap interval of either. The variance decompositions agree to within
$0.004$ in every $\eta^2$ (Table~\ref{tab:anova-decomp}).

The practical conclusion is that reproducibility must be claimed at the level at
which results are reported. Snapshot-pinned retrieval and seeded simulation make
the pipeline replayable, and the aggregate statistics are stable to within
$0.01\times$, but individual scenarios are not reproducible and should not be
cited as if they were. A supervisory deployment that relied on a specific named
scenario---rather than on a distribution over scenarios---would not survive a
re-run.

\subsection{Baselines, Crisis Envelopes and Calm-Period Comparators}
\label{sec:crisis-envelopes}

\begin{figure*}[t]
  \centering
  \includegraphics[width=0.95\linewidth]{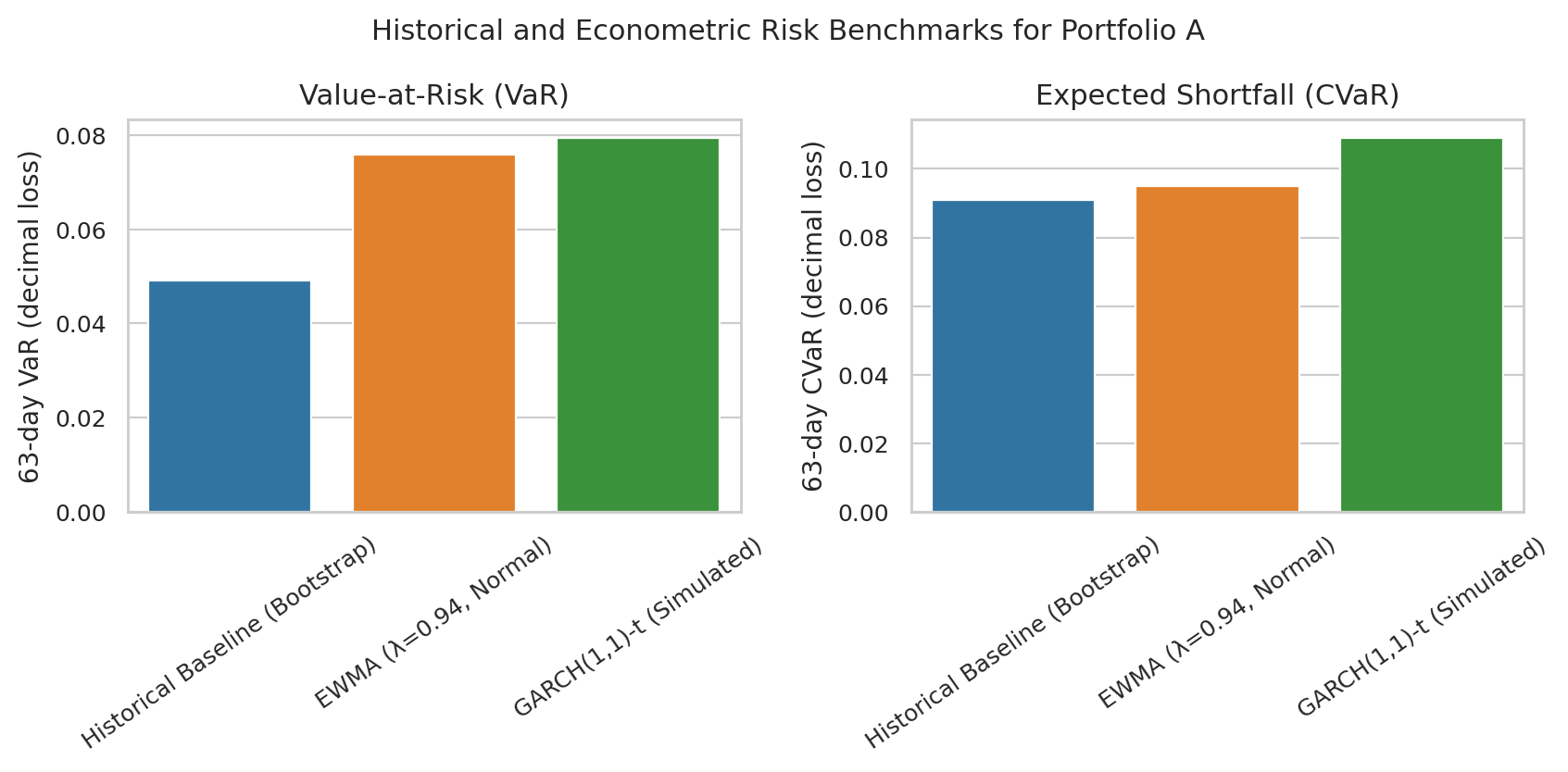}
  \caption{63-day VaR and CVaR (decimal losses) for Portfolio~A under the
  historical bootstrap, EWMA ($\lambda = 0.94$) and GARCH(1,1)--$t$ baselines.
  Numerical values appear in Table~\ref{tab:baseline-var}.}
  \label{fig:baselines}
\end{figure*}

Table~\ref{tab:baseline-var} reports the 63-day VaR and CVaR of Portfolio~A
under the historical bootstrap, EWMA and GARCH(1,1)--$t$ benchmarks. Portfolio~A
has a bootstrap VaR of $4.90\%$ and CVaR of $9.11\%$; Portfolio~B, $7.27\%$ and
$13.84\%$. Figure~\ref{fig:baselines} shows the same quantities graphically.

We note that risk does \textit{not} increase monotonically with model
flexibility. The bootstrap CVaR exceeds the EWMA CVaR because the historical
sample has a fatter left tail than the Normal assumption underlying the EWMA
rescaling, even though the EWMA VaR is higher. Earlier statements to the
contrary in this literature conflate the two metrics.

Two further caveats attach to these baselines and are stated here rather than
buried. First, the Portfolio~A bootstrap sample begins in November 2004, not
2000, because gold (GLD) has no earlier listed history; the sample window is
recorded in the artefact manifest. Second, the stress-period covariance matrix
$\Sigma_{\text{stress}}$ contains Global Financial Crisis observations only
through its pairwise half: the sector ETF XLRE lists only from October 2015, so
the complete-case half rests on 72 COVID-19 and 209 2022-inflation trading days.
Section~\ref{sec:experimental} describes the blend, and ablation S6 shows that
dropping the GFC altogether changes the pooled multiples by less than $0.2\%$.

\subsection{Extreme Scenarios}
\label{sec:results-extremes}

\begin{table*}[t]
  \centering
  \small
  \caption{Ten most severe accepted scenarios by linear-channel CVaR multiple
  (Portfolio~A, primary GPT-5.6-luna run). Shocks are deviations in
  percentage points from baseline. All ten originate from a single prompt
  variant.}
  \label{tab:top10}
  \begin{adjustbox}{max width=\textwidth}\begin{tabular}{llllrrrr}
    \toprule
    Country & RAG & News & Prompt variant & $\Delta g$ & $\Delta \pi$ & $\Delta r$ & CVaR mult. \\
    \midrule
    Italy          & Off & On  & v29\_em\_contagion & -7.30 & 2.54 & 2.50 & 3.61 \\
    Italy          & On  & On  & v29\_em\_contagion & -5.70 & 3.54 & 3.50 & 3.49 \\
    Italy          & On  & Off & v29\_em\_contagion & -6.70 & 2.44 & 2.50 & 3.45 \\
    Canada         & On  & On  & v29\_em\_contagion & -6.20 & 3.71 & 2.50 & 3.44 \\
    Canada         & Off & On  & v29\_em\_contagion & -6.20 & 2.71 & 2.50 & 3.42 \\
    Canada         & On  & Off & v29\_em\_contagion & -6.10 & 2.71 & 2.50 & 3.40 \\
    Germany        & On  & Off & v29\_em\_contagion & -6.90 & 2.57 & 1.50 & 3.38 \\
    France         & Off & On  & v29\_em\_contagion & -6.08 & 2.18 & 2.25 & 3.37 \\
    United States  & On  & On  & v29\_em\_contagion & -6.80 & 3.66 & 1.48 & 3.34 \\
    Germany        & On  & On  & v29\_em\_contagion & -6.80 & 1.45 & 1.50 & 3.34 \\
    \bottomrule
  \end{tabular}\end{adjustbox}
\end{table*}

Table~\ref{tab:top10} lists the ten most severe accepted scenarios by
linear-channel CVaR multiple for Portfolio~A. The tail is dominated by a single
prompt variant: all ten arise from \texttt{v29\_em\_contagion}, an
emerging-market currency and funding contagion narrative, spanning Italy,
Canada, Germany, France and the United States and appearing under three of the
four retrieval and news configurations.

That the extreme tail is a prompt artefact rather than a country or retrieval
artefact is consistent with the variance decomposition of
Section~\ref{sec:results-anova}, and reinforces the reading that severity in
this pipeline is inherited from scenario design.

This has a consequence for how every aggregate in this paper should be read. The
30 templates are manually designed experimental conditions, not a sample from
any characterised population of macroeconomic scenarios. Whether the reported
mean multiple is $4.1\times$ or something else depends on how many severe
narratives we chose to include, and the entire upper tail would disappear if we
had omitted a single template. The reported figures are therefore properties of
this prompt library under this translator, not estimates of an underlying stress
distribution, and we do not present them as the latter.

\subsection{Prompt Dispersion and Scenario Stability}
\label{sec:results-dispersion}

Table~\ref{tab:prompt-dispersion} reports, for each prompt and configuration,
the mean pairwise Euclidean distance between the seven countries' emitted
(growth, inflation, rate) levels. GPT exhibits dispersion of $1.55$--$1.64$ across both runs with
little variation across retrieval settings; Llama exhibits $2.73$--$3.09$,
roughly twice as much.
Intra-configuration dispersion within countries---the distance between
scenarios from \emph{different} prompts for the same country and
configuration---is $2.9$--$3.7$ for GPT across all configurations
(Table~\ref{tab:stability-config} shows the RAG-and-news cell) with tight
bootstrap intervals, the United States being the most dispersed and Japan the
least.

The contrast between prompt-level and configuration-level dispersion is
informative: scenarios generated from the \textit{same} prompt are tightly
clustered while scenarios generated from \textit{different} prompts within a
country are widely separated. This is the dispersion signature of prompt
anchoring documented in Section~\ref{sec:results-tiers}.

\subsection{Cross-Country Consistency Diagnostics}
\label{sec:results-fairness}

Table~\ref{tab:fairness} reports cross-country consistency diagnostics. We
compute these on individual accepted scenarios rather than on aggregated
factorial cells: with one scenario generated per
(country, prompt, RAG, news) cell, the two coincide except where a scenario was
rejected, and reporting a full $840$-cell grid when only $838$ scenarios were
accepted would overstate coverage. Coverage is therefore $838/840$, $838/840$
and $744/840$ for the three runs.

\begin{table*}[t]
  \centering
  \small
  \caption{Cross-country consistency diagnostics on accepted scenarios. Group
  gap is the max--min difference in mean VaR multiple across the seven
  economies. Outliers are counted on the linear-channel multiple by standard
  $z$-score and by a robust MAD-based score.}
  \label{tab:fairness}
  \begin{adjustbox}{max width=\textwidth}\begin{tabular}{llrrrrrr}
    \toprule
    Run & Pf. & $N$ & Mean & Gap (lin.) & Gap (nonlin.) & $|z|>3$ & $|z_{\text{MAD}}|>3.5$ \\
    \midrule
    \multirow{2}{*}{GPT (primary)}
      & A & 838 & 3.936 & 0.327 & 0.315 & 2 & 0 \\
      & B & 838 & 4.179 & 0.252 & 0.254 & 1 & 2 \\
    \midrule
    \multirow{2}{*}{GPT (repeat)}
      & A & 838 & 3.941 & 0.317 & 0.307 & 1 & 0 \\
      & B & 838 & 4.186 & 0.252 & 0.250 & 0 & 1 \\
    \midrule
    \multirow{2}{*}{Llama-3.1-8B}
      & A & 744 & 4.243 & 0.697 & 0.701 & 3 & 3 \\
      & B & 744 & 4.311 & 0.802 & 0.803 & 0 & 0 \\
    \bottomrule
  \end{tabular}\end{adjustbox}
\end{table*}

Group gaps in the two GPT runs are $0.25$--$0.33$, roughly $6$--$8\%$ of the
mean multiple, against $0.70$--$0.80$ for Llama, roughly $16$--$19\%$. Canada
is the most severe and Japan the least in both portfolios of both the primary
and the Llama run, which is the ordering the baseline regression of
Section~\ref{sec:results-country-artefact} predicts. Outliers are rare under
either criterion: at most three scenarios per portfolio in any run.

We deliberately avoid the term ``fairness'' for this analysis. There is no
protected attribute, no task-specific fairness criterion, and no normative
argument that equal risk multiples across the G7 would be desirable---different
baseline macro conditions are economically meaningful, and the seven economies
are not exchangeable observations. This is a cross-country consistency and
sensitivity check, and we name it accordingly.

\subsection{Variance Decomposition}
\label{sec:results-anova}

Table~\ref{tab:anova-decomp} reports partial $\eta^2$ from an ANOVA on
linear-channel VaR multiples, both pooled across portfolios and within
Portfolio~A. We report the within-portfolio decomposition as primary: the pooled
design places each scenario in the data twice, once per portfolio, sharing a
single LLM draw, so the residuals are not independent and the portfolio effect
is close to tautological given that the two portfolios have different baselines
and different covariance structures by construction.

\begin{table*}[t]
  \centering
  \small
  \caption{ANOVA variance decomposition of linear-channel VaR multiples, partial
  $\eta^2$. Within-portfolio figures (Portfolio~A) are primary; pooled figures
  are descriptive only.}
  \label{tab:anova-decomp}
  \begin{adjustbox}{max width=\textwidth}\begin{tabular}{llrrrr}
    \toprule
    & & \multicolumn{2}{c}{Pooled (A+B)} & \multicolumn{2}{c}{Portfolio A} \\
    \cmidrule(lr){3-4}\cmidrule(lr){5-6}
    Run & Effect & $p$ & $\eta^2$ & $p$ & $\eta^2$ \\
    \midrule
    \multirow{5}{*}{GPT (primary)}
      & portfolio      & 0.000 & 0.200 & --- & --- \\
      & country        & 0.000 & 0.129 & 0.000 & 0.140 \\
      & prompt variant & 0.000 & 0.865 & 0.000 & 0.873 \\
      & RAG            & 0.962 & 0.000 & 0.917 & 0.000 \\
      & news           & 0.827 & 0.000 & 0.649 & 0.000 \\
    \midrule
    \multirow{5}{*}{GPT (repeat)}
      & portfolio      & 0.000 & 0.204 & --- & --- \\
      & country        & 0.000 & 0.125 & 0.000 & 0.144 \\
      & prompt variant & 0.000 & 0.864 & 0.000 & 0.874 \\
      & RAG            & 0.825 & 0.000 & 0.677 & 0.000 \\
      & news           & 0.080 & 0.002 & 0.101 & 0.003 \\
    \midrule
    \multirow{5}{*}{Llama-3.1-8B}
      & portfolio      & 0.078 & 0.002 & --- & --- \\
      & country        & 0.000 & 0.109 & 0.000 & 0.087 \\
      & prompt variant & 0.000 & 0.390 & 0.000 & 0.398 \\
      & RAG            & 0.272 & 0.001 & 0.374 & 0.001 \\
      & news           & 0.000 & 0.010 & 0.020 & 0.008 \\
    \bottomrule
  \end{tabular}\end{adjustbox}
\end{table*}

Prompt variant dominates in every specification, with
$\eta^2 \approx 0.87$ for GPT and $0.40$ for Llama. Two qualifications are
essential and we state them before any interpretation.

First, the figure is inflated by design. Severity tiers are assigned at the
template level, so template identity carries a deliberate severity signal by
construction. An $\eta^2$ of $0.87$ is close to the statistical signature of
having told the model how severe the scenario should be and then measured
whether the output was severe. It is not evidence of an emergent model property.

Second, the cross-model comparison of $\eta^2$ is confounded. Partial $\eta^2$
carries residual variance in its denominator, and Llama's within-tier dispersion
is four to seven times GPT's, which mechanically depresses its $\eta^2$ even
if the absolute prompt effect were identical. Comparing absolute contrasts
instead reverses the apparent conclusion: the between-tier spread in $\Delta g$
is \emph{larger} for GPT ($4.35$~pp, $95\%$ CI $[4.24, 4.44]$) than for Llama
($2.59$~pp, $95\%$ CI $[2.02, 3.12]$). We therefore withdraw the inference that
Llama's lower $\eta^2$ measures how much of the GPT effect is prompt compliance;
the statistics do not support it. What the absolute contrasts do show is that
both models order the tiers as instructed, and that GPT does so with far less
residual spread.

Country effects are modest ($\eta^2 = 0.09$--$0.14$ within Portfolio~A) and, per
Section~\ref{sec:results-country-artefact}, largely reflect baseline dispersion.
Retrieval is null in every run. News is null for both GPT runs but small and
statistically significant for Llama ($\eta^2 = 0.008$, $p = 0.02$ within
Portfolio~A; $\eta^2 = 0.010$, $p < 0.001$ pooled). The repeat run reproduces
every $\eta^2$ of the primary run to within $0.004$. Within-portfolio and pooled
decompositions tell the same story, and the pooled portfolio effect
($\eta^2 = 0.20$ for GPT, $0.002$ for Llama) is close to tautological for the
reason given above.

\paragraph{Dependence structure and cluster-robust inference.} The ANOVA above
assumes independent residuals, which this design does not deliver: the same 30
prompt templates recur across all seven countries, each LLM draw is reused
across two portfolios, and every scenario's VaR is itself a Monte Carlo estimate
from a shared set of PCA loadings and covariance matrices. Ordinary $p$-values
are therefore anti-conservative. We re-estimate the linear-channel model on
Portfolio~A alone with standard errors clustered on \texttt{prompt\_variant},
the generation unit shared across countries
(Table~\ref{tab:cluster-robust}). The substantive conclusions are unchanged. For GPT the retrieval and news
coefficients are within $0.03$ multiples of zero and insignificant
($p \geq 0.26$). For Llama the news coefficient is $-0.14$ (robust SE $0.07$,
$p = 0.04$): the small news effect of Table~\ref{tab:anova-decomp} survives
clustering, but only marginally. Without severity bands (P4), GPT's news
coefficient is $+0.05$ ($p = 0.02$).

\begin{table*}[t]
  \centering
  \small
  \caption{Cluster-robust inference on the linear-channel VaR multiple,
  Portfolio~A only, standard errors clustered on prompt variant
  (30 clusters). Country is tested jointly across its six contrasts.}
  \label{tab:cluster-robust}
  \begin{adjustbox}{max width=\textwidth}\begin{tabular}{llrrr}
    \toprule
    Run & Effect & Coefficient & Robust SE & $p$ \\
    \midrule
    \multirow{3}{*}{GPT (primary)}
      & RAG      & $+0.0018$ & 0.0159 & 0.908 \\
      & News     & $+0.0080$ & 0.0215 & 0.709 \\
      & Country (joint) & --- & --- & $<10^{-17}$ \\
    \midrule
    \multirow{3}{*}{GPT (repeat)}
      & RAG      & $+0.0072$ & 0.0131 & 0.581 \\
      & News     & $+0.0285$ & 0.0252 & 0.257 \\
      & Country (joint) & --- & --- & $<10^{-9}$ \\
    \midrule
    \multirow{3}{*}{Llama-3.1-8B}
      & RAG      & $+0.0525$ & 0.0630 & 0.405 \\
      & News     & $-0.1371$ & 0.0674 & 0.042 \\
      & Country (joint) & --- & --- & $<10^{-19}$ \\
    \midrule
    \multirow{3}{*}{GPT (unanchored, P4)}
      & RAG      & $+0.0116$ & 0.0103 & 0.260 \\
      & News     & $+0.0507$ & 0.0218 & 0.020 \\
      & Country (joint) & --- & --- & $<10^{-13}$ \\
    \bottomrule
  \end{tabular}\end{adjustbox}
\end{table*}

We do not report a cluster-robust test for prompt variant itself, since prompt
variant defines the clusters.

Applying the same decomposition to the raw macro shocks rather than the risk
multiples gives an even starker picture: prompt variant reaches
$\eta^2 = 0.950$ on $\Delta g$ for GPT against $0.413$ for Llama, while country
reaches $0.299$ and $0.188$ respectively. On inflation, where no magnitude band
is prescribed, prompt variant still explains $\eta^2 = 0.916$ for GPT but only
$0.637$ for Llama, and retrieval attains marginal significance for Llama alone
($p = 0.006$, $\eta^2 = 0.011$). News moves Llama's growth shock directly
($p = 0.001$, $\eta^2 = 0.015$), which is the route by which it reaches the
risk layer; for GPT its largest effect is on inflation in the repeat run
($\eta^2 = 0.012$), too small to register in the multiples.

\begin{figure*}[t]
  \centering
  \includegraphics[width=\linewidth]{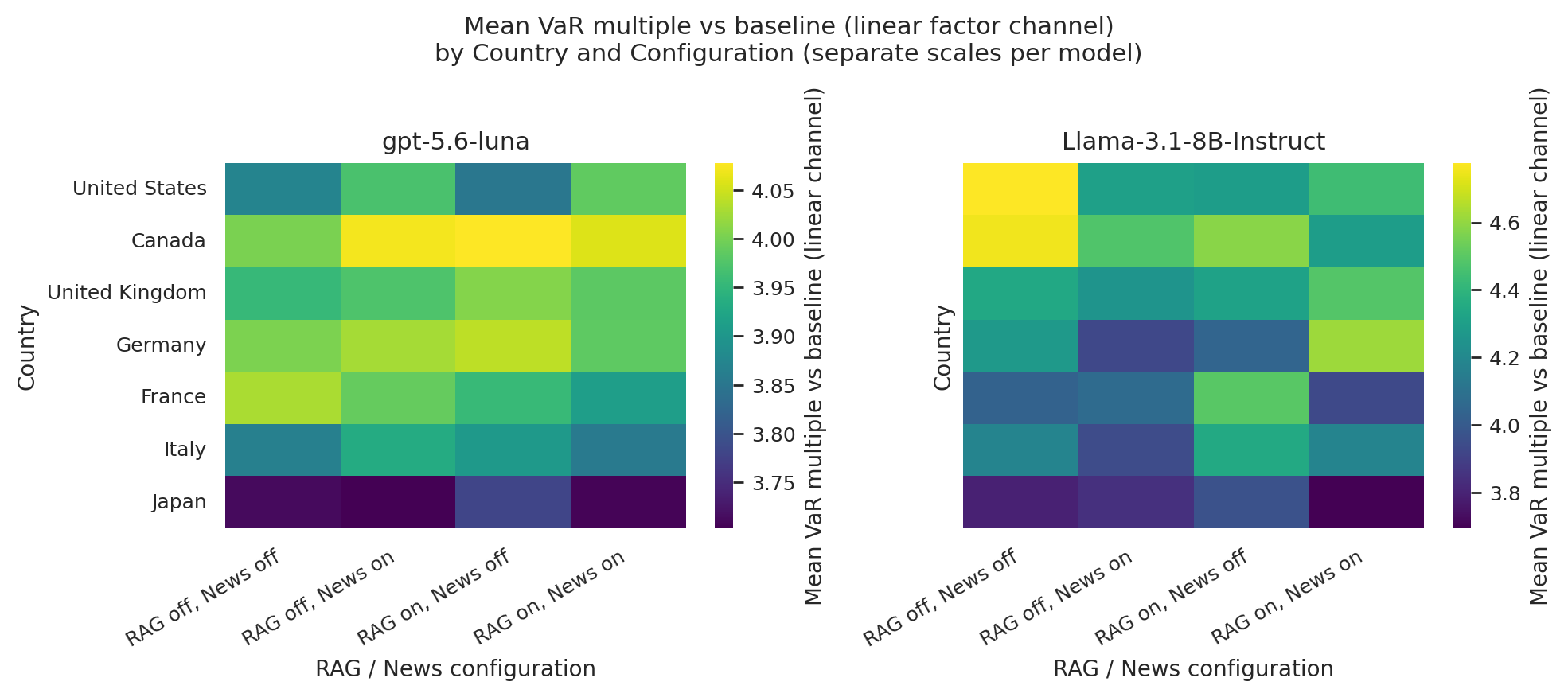}
  \caption{Heatmaps of mean linear-channel VaR multiples for Portfolio~A by
  country (rows) and retrieval/news configuration (columns), shown separately by
  model.}
  \label{fig:heatmaps}
\end{figure*}

Figure~\ref{fig:heatmaps} displays the mean linear-channel VaR multiple by
country and configuration for each model. Cross-country variation is visible and
systematic, while movement across retrieval and news columns is not.

\subsection{Ablations}
\label{sec:results-ablations}

Table~\ref{tab:ablation-results} collects the twelve presets of
Table~\ref{tab:ablations} on common metrics: pooled linear- and
volatility-channel multiples and the Portfolio~A variance decomposition of the
linear VaR multiple. Five readings follow.

\begin{table*}[t]
  \centering
  \small
  \caption{Ablation results. Multiples are pooled over both portfolios and
  expressed relative to the historical bootstrap baseline. $\eta^2$ and the
  news $p$-value are from the Portfolio~A ANOVA of the linear VaR multiple;
  retrieval is null in every run ($p \geq 0.35$). S3 reproduces P1 exactly
  because the empirical $\kappa$ is unusable.}
  \label{tab:ablation-results}
  \begin{tabular}{llrrrrrrrr}
    \toprule
    & & & \multicolumn{2}{c}{Linear} & \multicolumn{2}{c}{Volatility} & \multicolumn{3}{c}{Portfolio~A ANOVA} \\
    \cmidrule(lr){4-5}\cmidrule(lr){6-7}\cmidrule(lr){8-10}
    Preset & Change & $N$ & VaR & CVaR & VaR & CVaR & $\eta^2_{\text{prompt}}$ & $\eta^2_{\text{country}}$ & $p_{\text{news}}$ \\
    \midrule
    P1 & primary              & 838 & 4.058 & 2.438 & 2.816 & 1.824 & 0.873 & 0.140 & 0.649 \\
    P2 & repeat draw          & 838 & 4.064 & 2.442 & 2.823 & 1.828 & 0.874 & 0.144 & 0.101 \\
    P3 & Llama                & 744 & 4.277 & 2.560 & 2.916 & 1.886 & 0.398 & 0.087 & 0.020 \\
    P4 & no severity bands    & 840 & 3.355 & 2.062 & 2.625 & 1.704 & 0.638 & 0.197 & 0.002 \\
    \midrule
    S1 & $\kappa = 2\%$       & 838 & 3.481 & 2.144 & 2.816 & 1.824 & 0.872 & 0.141 & 0.673 \\
    S2 & $\kappa = 5\%$       & 838 & 4.605 & 2.716 & 2.816 & 1.824 & 0.873 & 0.138 & 0.646 \\
    S3 & empirical $\kappa$   & 838 & 4.058 & 2.438 & 2.816 & 1.824 & 0.873 & 0.140 & 0.649 \\
    S4 & rate base $-1$~pp    & 838 & 4.310 & 2.570 & 2.844 & 1.841 & 0.871 & 0.136 & 0.661 \\
    S5 & rate base $+1$~pp    & 838 & 3.810 & 2.309 & 2.800 & 1.814 & 0.874 & 0.138 & 0.672 \\
    S6 & complete-case $\Sigma$ & 838 & 4.060 & 2.441 & 2.818 & 1.827 & 0.873 & 0.140 & 0.643 \\
    S7 & core PCA universe    & 838 & 4.073 & 2.447 & 2.816 & 1.824 & 0.919 & 0.244 & 0.790 \\
    S8 & Llama, stated shocks & 780 & 3.994 & 2.411 & 2.870 & 1.857 & 0.361 & 0.013 & 0.045 \\
    \bottomrule
  \end{tabular}
\end{table*}

\textbf{Without severity bands, GPT writes a much milder recession.} P4 is the
disambiguating experiment: identical to P1 except that templates prescribe no growth-shortfall band. The mean growth deviation rises from $-3.85$ to $-2.20$~pp, a shortfall
$1.6$~pp smaller, and its standard deviation falls from $1.48$ to $0.63$~pp.
GPT labels $95\%$ of these scenarios \emph{severe} and none \emph{contained}
(28 of the 30 templates are modally severe),
and the mean soft plausibility score falls from $4.66$ to $4.03$. The linear-channel multiples fall by $17\%$ (VaR) and
$15\%$ (CVaR). Prompt variant still dominates but its share falls from
$\eta^2 = 0.87$ to $0.64$, country rises from $0.14$ to $0.20$, and news becomes
significant ($\eta^2 = 0.012$, $p = 0.002$). The severe tail of the tiered corpus
is therefore supplied by the bands, not by the model; left to itself GPT
converges on a shallow, compressed downturn, and it is only then that its
context begins to move the numbers. Narrative--numeric coupling does not improve
(pooled $r = +0.014$).

\textbf{Shock source removes the country effect.} Using Llama's stated changes
(S8) rather than level-minus-baseline deviations eliminates the country effect
and closes most of the Llama--GPT severity gap
(Sections~\ref{sec:results-country-artefact} and~\ref{sec:results-risk}).

\textbf{Calibration and the rate baseline move the level, not the structure.}
$\kappa$ and a $\pm 1$~pp rate-baseline shift move the linear multiples by
$-14\%$ to $+13\%$ and $\pm 6\%$ respectively, but leave every $\eta^2$ within
$0.005$ of P1 and the volatility channel within $1\%$.

\textbf{The stress-covariance estimator is immaterial.} Excluding the GFC from
$\Sigma_{\text{stress}}$ (S6) changes no pooled multiple by more than
$0.2\%$ and no $\eta^2$ by more than $0.001$.

\textbf{The factor-extraction universe barely affects the linear channel.}
Extracting the factors from the three Portfolio~A assets alone (S7), which makes
the curvature channel identically zero for that portfolio, moves the pooled
linear multiples by under $0.4\%$ but raises both the prompt and the country
share of Portfolio~A variance ($0.92$ and $0.24$), consistent with the factors
now spanning the portfolio exactly and leaving less residual variation.

\subsection{Sector Attribution}
\label{sec:results-sectors}

Scenarios emit on average $4.03$ sector tags each for GPT and $3.19$ for Llama.
The raw label space is large---409 and 237 distinct strings respectively---which
is why canonicalisation to the traded ETF taxonomy is required before
aggregation. The modal attributions are economically sensible and stable across
the two GPT runs: banks ($5.8\%$ of mentions), commercial real estate ($5.2\%$),
construction ($4.5\%$), transportation ($3.3\%$), and utilities and materials
($2.5\%$ each). The repeat run shares all ten of the top ten labels (rank
correlation $0.99$). Llama's distribution is flatter and differently ordered, led
by manufacturing ($8.0\%$), banks ($4.9\%$) and construction ($4.8\%$).

Sector attribution is the most \emph{stable} narrative output we measure: the
two GPT runs agree closely on the ranking and frequency of the top ten sectors
despite differing on $43\%$ of scenario titles. Stability is not accuracy. We
have no external benchmark against which these attributions could be scored, so
we claim only that they are reproducible across repeat draws, not that they are
correct or economically appropriate.

\section{Discussion}
\label{sec:discussion}

This section interprets the findings of Section~\ref{sec:results} and draws
implications for LLM-driven macro--financial stress testing. Our reading is
deliberately conservative: several results qualify rather than support the
optimistic framing common in this literature, and we think the qualifications
are the more useful contribution.

\subsection{Where the Model's Contribution Actually Lies}
\label{sec:disc-contribution}

The dominant empirical pattern across all three runs is that quantitative
severity is inherited from the prompt. Prompt variant explains
$\eta^2 \approx 0.87$ of the variance in tail-risk multiples for GPT-5.6-luna
and $\eta^2 = 0.950$ of the variance in the raw growth shock; within-tier
dispersion is a fraction of the width of the prescribed magnitude band; and the
ten most extreme scenarios all originate from a single prompt variant. Apparent
country heterogeneity is $87$--$91\%$ explained by regressing on the baseline
growth rate ($R^2 = 0.843$, $0.820$, $0.796$), and it vanishes for the
open-weight model once its own stated changes are used (S8). Retrieval moves
nothing; news moves the result only where the prompt leaves magnitude free.

It would be possible to present this as a positive finding about the importance
of prompt engineering. We think that reading is misleading, and we go further:
because the severity bands are supplied in the prompt, this result is largely
\emph{designed in} and should be treated as a diagnostic observation rather than
as a discovery. The honest summary is that, in this configuration, a capable
instruction-following model returns approximately the number it was asked for.
The disambiguating experiment---removing the growth-shortfall band from every template (P4)---shows what GPT produces without the anchor: a recession
$1.6$~pp shallower on average and less than half as dispersed, multiples
$15$--$17\%$ lower, and a prompt share of $0.64$ rather than $0.87$
(Section~\ref{sec:results-ablations}). The severe tail of the tiered corpus is
supplied by the bands; the model's own calibration, in this configuration, is a
shallow and compressed downturn.

This does not make the approach useless, but it relocates its value. Three
outputs remain genuinely model-generated and are not prescribed by our
templates: the transmission narrative, the sector attribution, and the joint
response of inflation and the policy rate conditional on the growth path. The
last of these is two-sided and substantially dispersed
(Table~\ref{tab:macro-summary}), and the sector attributions are stable across
repeat runs even when the underlying text is not
(Section~\ref{sec:results-sectors}). We therefore position the pipeline as an
elicitation tool for narrative and exposure structure operating \textit{around}
a prescribed severity path---a role closely analogous to how CCAR and EBA
exercises supply severity and leave institutions to reason about transmission.

\subsection{Retrieval as Narrative Grounding, Not Risk Modulation}
\label{sec:disc-rag}

Retrieval has no detectable effect on tail risk in either factor channel in any
run, and news has none for GPT under tiered bands. Rather than treat this as a general result about retrieval-augmented
generation, we attribute it to specific and identifiable architectural
decisions, and we report them so that the null can be interpreted.

First, the retrieval corpus is intentionally narrow: country profiles share a
common WEO template and differ chiefly in three numbers, so nearest neighbours
are economically similar peers with overlapping content. Second, the prompt
template constrains both the output format and the severity range, so retrieved
context acts as a topical anchor rather than a driver of magnitude. Third, and
decisively, text enters the risk translator \textit{only} through the curvature
term of the nonlinear channel, and that term contributes under $0.4\%$ of the
drift (Section~\ref{sec:results-risk}). A signal routed exclusively through a
channel of that magnitude cannot move the result, whatever its content. The only
other route is through the emitted numbers, and the ablations show it opening
exactly where expected: news has a small, significant effect for Llama, which
follows the prompted bands less tightly, and for GPT once the bands are removed.

The conclusion we draw is architectural rather than empirical: if a pipeline is
intended to test whether grounding affects quantitative risk, the grounding must
enter somewhere consequential. In our design the natural candidate is the regime
severity index $\lambda_s$, which already reads the narrative, although in the
current runs retrieval does not move it detectably (volatility-channel $p = 0.56$
for the primary run). We did not restructure the
pipeline to exploit this, because doing so post hoc would amount to selecting an
architecture on the basis of the significance it produces.

\subsection{Narrative and Numeric Severity Are Weakly Coupled}
\label{sec:disc-coupling}

Section~\ref{sec:results-grounding} shows that within the severe and systemic
tiers---nearly three quarters of the corpus---the association between the
realised growth deviation and the narrative severity score is indistinguishable
from zero even after correcting for range restriction, and that the open-weight
model shows no association in any tier under negligible restriction. The one
exception is the contained tier under GPT, where correcting for a
$5.8\times$ restriction reveals a strong association ($r_{\text{corr}} = -0.85$).
Removing the bands does not help: the unanchored run shows no pooled association
at all ($r = +0.014$).
Outside that tier, a scenario can read as a catastrophe and carry a mild number,
or the reverse, without any internal mechanism flagging the mismatch.

We are careful about what this licenses. The instrument is an off-the-shelf NLI
classifier with no validation against expert judgement, and measurement error
attenuates correlations toward zero, so the evidence is consistent with a weaker
true coupling than we would like to detect rather than with none at all.

This is a substantive limitation for supervisory use, because the two artefacts
serve different audiences. The narrative is what a risk committee reads and
debates; the number is what propagates into capital planning. If they are
decoupled, human review of the narrative provides weaker assurance over the
quantitative output than it appears to.

It also has a concrete consequence inside our own architecture, noted in
Section~\ref{sec:results-grounding}: $\lambda_s$ and the factor drift are driven
by near-independent signals, so the volatility channel and the factor channels
do not represent decompositions of a single severity concept. We report the
channels separately for this reason and caution against summing or averaging
them.

\subsection{Model Behaviour: An Instruction-Following/Diversity Trade-off}
\label{sec:disc-models}

On identical templates, identical retrieval artefacts and an identical news
snapshot, the two model families fail in opposite directions.

GPT-5.6-luna is arithmetically exact: in $100\%$ of scenarios its stated change
versus baseline reconciles with the level it reports, with a median discrepancy
of $0.00$~pp. But it anchors tightly on the prescribed severity band, with
within-tier standard deviations of $0.25$--$0.60$~pp and tier means within
$0.2$~pp of the band midpoints.

Llama-3.1-8B-Instruct inverts this. Its within-tier dispersion is four to seven
times larger, its tier labels are less consistent within a template, and it re-labels severity more
often---all indicators of genuine scenario diversity. Yet in $89.5\%$ of
accepted scenarios its stated change contradicts the level it reports by more
than $0.5$~pp in at least one variable, with a median discrepancy of
$1.70$~pp. When the stated change
rather than the level is propagated (S8), its country effect disappears and its
tail-risk multiples fall to GPT's level. Its least severe accepted scenario sits exactly at
the $-0.5$~pp deviation floor, so the gate is binding on it in a way it is not
for GPT. It is not grounding itself in the baseline it was given.

Neither behaviour is straightforwardly preferable. A stress-testing pipeline
needs both arithmetic reliability and scenario diversity, and neither model
supplies both. We note that the failure modes are invisible from the narratives
alone: reading either corpus, an analyst would not detect that one model has
collapsed its severity distribution onto the prompt and the other has lost track
of its baseline. Both were caught only by automated cross-checks
(the stated-versus-computed reconciliation and the dispersion diagnostic), which
we accordingly recommend as standard instrumentation rather than optional
diagnostics.

\subsection{Reproducibility at the Right Level of Aggregation}
\label{sec:disc-repro}

Our two GPT runs differ on $43\%$ of scenario titles and roughly $45\%$ of
individual shock values, yet agree on the pooled linear VaR multiple to
$0.006\times$ and on every variance share to $0.004$. Both statements are true and they matter for different claims.

The literature's usual formulation---that low-temperature decoding yields
reproducible outputs---is not supported. The GPT-5 family accepts only
$\text{temperature}=1$, so greedy decoding is unavailable, and provider-side
batching and kernel selection introduce variation that no client-side seed
controls. What \textit{is} reproducible is the aggregate: given frozen
retrieval artefacts, a pinned news snapshot and a seeded simulation engine,
distributional statistics replicate to within $0.01\times$.

We therefore recommend that pipelines of this kind state reproducibility at the
level of the reported statistic, and that individual scenarios never be cited as
stable objects. A supervisory process built around a named scenario would not
survive re-execution; a process built around a scenario distribution would.

We also document two infrastructure failure modes that silently break
reproducibility and are easy to miss. Embedding on GPU rather than CPU perturbs
MiniLM vectors at the $10^{-7}$ level, which propagates chaotically through the
$k$-means selection of headline exemplars and changes the prompts themselves; we
measured four of twenty exemplars changing under a $10^{-7}$ perturbation.
Separately, re-querying the news API between runs returns different headlines
for a nominally identical window, because the provider returns the most recent
articles within that window and the set grows over time. Both are addressed by
pinning---CPU-only embedding and a frozen, hash-recorded headline
snapshot---and both would otherwise have confounded the cross-model comparison
with a change in inputs.

\subsection{Limitations}
\label{sec:disc-limitations}

Beyond the substantive findings above, several constraints bound what this study
can claim.

\textbf{Severity is prescribed, not discovered.} The tiered magnitude bands in
the prompt templates make the growth shock distribution partly a design choice.
We report this rather than tune it away. The unanchored run (P4) shows what
GPT does without the anchor---a much shallower and more compressed downturn---but
it is a single run of a single model, and it cannot say which of the two
distributions is the more \emph{plausible}; that requires external validation.

\textbf{The crisis covariance is only partly a GFC covariance.} Because XLRE
lists only from 2015, the GFC enters $\Sigma_{\text{stress}}$ only through the
pairwise half of the blend, and a PSD repair is needed to make that half valid.
Excluding the GFC entirely (S6) changes the results by less than $0.2\%$, so the
choice is immaterial here, but the volatility channel should not be read as a
calibrated GFC stress.

\textbf{Headline severity is a calibration choice.} The reported multiples scale
linearly with $\kappa$, and the proximity of the central calibration to the GFC
envelope is a consequence of setting $\kappa = 3.5\%$. We report the full
interval for this reason.

\textbf{Factor identification is imperfect.} Under the balanced extraction
universe, PC1 anchors strongly on equities ($r = 0.95$) and PC3 on duration
($r = 0.83$), but PC2 anchors only moderately on gold ($r = 0.58$), and PC3
loads comparably on gold and Treasuries. We describe the factors by their
loadings rather than as pure equity, inflation and rates factors, and report the
full anchor-correlation matrix.

\textbf{The regime classifier is an unvalidated instrument.} It never assigns
the \texttt{normal} label, compressing the effective severity score to
$[0.49, 0.99]$, and it has never been benchmarked against human judgements of
macro-financial severity. Dispersion is not validity. Attenuation from
measurement error biases the reported correlations toward zero, so our
narrative-numeric result should be read as a statement about DeBERTa-derived
scores rather than about narrative severity as an expert would assess it.

\textbf{The macro-to-factor mapping is stipulated, not estimated.}
Equation~\ref{eq:factor-shock} is a definition, and $\kappa$ is set by hand.
Our attempt to estimate it (S3) on 43 quarters of US data produced a wrong-signed
equity elasticity and a $\kappa$ interval that contains zero, so there is still no
usable estimated relationship between a one percentage point growth deviation
and any number of factor standard deviations, and no out-of-sample validation of
the transmission. Every reported multiple is consequently a
hand-set constant applied to a prompt-set deviation. The remedy is known---regress
the extracted factors on realised macro surprises from forecast revisions or
consensus surprise series and use estimated coefficients with standard
errors---on a longer, multi-country sample---and we identify it as the priority
extension in Section~\ref{sec:disc-future}. Until it is done, the multiples in this paper are
conditional stress scores under a stated translator, not measurements of
portfolio risk.

\textbf{The reported tail is a shifted mean, not a fattened tail.} As
Section~\ref{sec:results-risk} shows, the scenario distributions have
$\text{CVaR}/\text{VaR} \approx 1.15$ against a baseline ratio of $1.86$.

\textbf{Two runs do not characterise reproducibility.} The repeat-run agreement
we report establishes that aggregate stability exists, not its sampling
distribution. More repeat runs would be needed to say how reliably it holds.

\textbf{Two models, one language, one horizon.} We evaluate one proprietary and
one open-weight model, English-language retrieval only, and a single
five-quarter-ahead horizon. The instruction-anchoring behaviour we document for
GPT may not generalise across model families or prompt styles.

\textbf{The plausibility gate is weakly binding.} It rejects $0.2\%$ of GPT
scenarios and none at all once the severity bands are removed. On these runs it is better described as a guardrail against malformed
or internally contradictory output than as a filter that shapes the accepted
distribution.

\subsection{Ethics and Societal Impact}
\label{sec:disc-ethics}

Three risks specific to this application deserve statement.

\textbf{Misplaced confidence in narrative review.} Because narrative and numeric
severity are weakly coupled (Section~\ref{sec:disc-coupling}), human review of
scenario prose gives less assurance over the quantitative output than it
intuitively appears to. Institutions adopting such pipelines should review the
numbers directly and instrument the coupling.

\textbf{Systemic homogenisation.} If many institutions adopt similarly prompted
models, and severity is predominantly inherited from the prompt as we find, then
scenario libraries across the system may converge far more than the diversity of
the narratives suggests. Prompt diversity, model diversity and expert
adjustment are the relevant safeguards.

\textbf{Data provenance.} The pipeline uses only public macroeconomic
projections and public news headlines, snapshotted with recorded hashes, and
retains no personal data. We note that free-tier news APIs commonly restrict use
to development rather than publication, and institutional deployments must
verify licensing.

\subsection{Implications and Future Work}
\label{sec:disc-future}

Four directions follow directly from the findings.

\textbf{Estimate the transmission empirically.} The single change that would
most improve this work is to replace the stipulated macro-to-factor mapping with
an estimated one---via local projections, a structural VAR, a dynamic factor
model, or a macro-conditioned return model, on a longer and multi-country sample
than our unsuccessful 43-quarter US attempt---reporting elasticities with
standard errors, an out-of-sample validation, and sensitivity to alternative
transmission specifications. Only then can the pipeline's output be described as
a risk estimate rather than a conditional stress score.

\textbf{Validate unanchored severity.} Without a band, GPT produces a shallow and
compressed downturn; with one, it reproduces the band. Neither run tells us which
is closer to a plausible stress. A design that elicits the transmission mechanism
first and the magnitude only as a derived consequence, scored against expert
ratings or realised outcomes, is the natural next experiment, and it should be
repeated across model families.

\textbf{Route grounding through a consequential channel.} Retrieval should enter
where it can matter, most plausibly through the regime index rather than a
capped curvature term, and the effect should be pre-registered rather than
discovered.

\textbf{External validation.} No result here establishes that these scenarios
are \textit{good}, only that they are internally coherent and reproducible in
aggregate. A retrospective exercise---generating scenarios for a horizon that
has since elapsed and scoring them against realised outcomes---or a structured
expert-rating study would address the gap that plausibility gating cannot.

\textbf{Richer translators, with commensurate governance.} Multi-regime
stochastic volatility, credit and FX channels, and multi-country VAR systems
would capture dynamics the present translator omits. We deliberately retained a
simple, auditable translator so that the LLM's contribution could be isolated;
any extension should be paired with the diagnostic instrumentation this study
found necessary.

\section{Conclusion}
\label{sec:conclusion}

This paper presented a transparent, auditable pipeline that couples
retrieval-augmented large language models to a deterministic macro--financial
risk translator, and evaluated it across 2{,}420 accepted scenarios drawn from
three runs of two model families over the G7, together with an unanchored-severity
run and eight risk-layer ablations. The two GPT runs share an
identical experimental configuration and are repeat draws rather than
independent designs; only the model family distinguishes the third. Country fundamentals
from the April-2026 IMF \textit{World Economic Outlook}, optionally enriched
with a frozen news snapshot, are embedded with MiniLM and indexed with FAISS.
GPT-5.6-luna and Llama-3.1-8B-Instruct emit structured JSON scenarios for a
Q4--2027 horizon; the emitted macroeconomic \textit{levels} are converted to
deviations from the institutional baseline and mapped into two stylised
portfolios through a three-channel PCA translator with regime-mixed covariance.

Our headline quantitative result is that LLM-generated scenarios imply
linear-channel tail-risk multiples of $3.5$--$4.6\times$ VaR and
$2.1$--$2.7\times$ CVaR for GPT-5.6-luna, and $3.7$--$4.9\times$ and
$2.2$--$2.9\times$ for Llama-3.1-8B, where the intervals span a macro-equity
sensitivity assumption of $2\%$ to $5\%$ per percentage point of growth
shortfall. We report intervals rather than point estimates because the response
is linear in that assumption ($R^2 = 0.9998$), the level of every multiple
depends on it, and our attempt to estimate it empirically failed.

The more consequential findings are qualifications, and two of them are
qualifications of our own design. Severity is inherited from scenario design
rather than generated by the model: removing our severity bands yields a recession
$1.6$~pp shallower and multiples $15$--$17\%$ lower. Retrieval is inert and news
nearly so, but both can reach the risk layer only through the emitted numbers, and
news does so measurably only where the prompt leaves magnitude free---so this too
follows from a design choice rather than from evidence about grounding. The
third finding is empirical: after correcting for range restriction, the
association between narrative and numeric severity remains indistinguishable
from zero in the severe and systemic tiers and, for the open-weight model,
throughout---so human review of scenario prose provides weaker assurance over
the quantitative output than it appears to. The fourth is a behavioural contrast
on identical prompts: GPT reconciles its own baseline arithmetic in every
scenario but anchors on the prompted magnitude band, whereas Llama generates
four to seven times wider within-tier dispersion while contradicting its stated
changes in $90\%$ of cases---and the apparent country structure and extra
severity of its scenarios largely disappear once its stated changes are used.

On reproducibility we find that the level of aggregation determines the claim.
Two runs under identical settings share only $54\%$ of individual growth shocks
and $57\%$ of scenario titles, yet agree on the pooled linear VaR multiple to
$0.006\times$. Deterministic decoding is unavailable for the GPT-5 family, so
claims of deterministic generation would be false; what the pipeline supports is
snapshot-replayable aggregate statistics. We further document two infrastructure
failure modes---GPU-versus-CPU embedding perturbing prompt construction through
chaotic $k$-means exemplar selection, and news-API re-querying returning
different headlines for a nominally fixed window---that silently break
comparability and are addressed here by CPU-pinned embedding and hash-recorded
headline snapshots.

Taken together, these results suggest a narrower but more defensible role for
LLMs in supervisory stress testing than the literature typically claims. They
are not autonomous calibrators of stress magnitude: in our setting the magnitude
is what the prompt asked for. They are, however, effective elicitation tools for
transmission narratives, sector exposure maps and the joint inflation and policy
response conditional on a prescribed growth path---and sector attribution in
particular proved the most stable output we measured, agreeing closely across
repeat runs that differed on $43\%$ of scenario titles. Positioned that way,
around a severity path supplied by a regulator or a risk committee rather than
by the model, and instrumented with the arithmetic, dispersion and grounding
diagnostics developed here, LLM scenario generation is a credible complement to
established frameworks. Positioned as a replacement for scenario calibration, it
is not.



\bibliographystyle{ACM-Reference-Format}

\bibliography{cas-refs}



\clearpage
  \appendix

\section{Scenario Example}
\label{app:scenario-example}

The excerpt below illustrates a typical accepted scenario under the unified JSON
template (GPT-5.6-luna, RAG on, news on; Canada, Q4--2027), reproduced verbatim
from the run artefacts. Note that \texttt{gdp\_growth}, \texttt{inflation} and
\texttt{interest\_rate} are horizon \emph{levels}; the \texttt{stated\_*\_change}
fields are the model's own reported deviations, and \texttt{d\_*} are the
deviations computed from level minus baseline. The \texttt{gap\_*} fields record
the reconciliation between the two, which for this scenario agrees to within
$0.005$~pp.

\begin{lstlisting}[basicstyle=\ttfamily\footnotesize,breaklines]
{
  "country": "Canada",
  "title": "Financial Contagion Scenario",
  "gdp_growth": -3.8,
  "inflation": 1.4,
  "interest_rate": 1.25,
  "rationale": "A disorderly repricing of U.S. corporate credit and
    technology equities triggers a broad global risk-off episode, with
    leveraged funds and banks facing losses and tighter wholesale
    funding. Canadian financial institutions experience wider risk
    premia and reduced market liquidity, while a weaker Canadian dollar
    raises imported costs but collapsing commodity prices, housing
    wealth, and U.S. demand depress domestic spending and exports. The
    Bank of Canada supplies liquidity, eases the policy rate, and
    activates macroprudential flexibility, but credit tightening and
    confidence effects persist through Q4 2027. Real GDP growth falls
    to -3.8%, disinflation from the demand shock outweighs
    exchange-rate pass-through, inflation declines to 1.4%, and the
    policy rate falls to 1.25%.",
  "risk_sectors": ["banks", "capital markets", "housing",
                   "energy exporters", "high-yield corporates"],
  "stated_gdp_growth_change": -5.7,
  "stated_inflation_change": -0.70,
  "stated_interest_rate_change": -1.0,
  "severity_tier": "systemic",
  "rag": true,
  "use_news": true,
  "model": "gpt-5.6-luna",
  "provider": "OpenAI",
  "prompt_variant": "v10_contagion",
  "prompt_template_version": "1d7e6dee3dc7",
  "prompt_hash": "f3c57f5064926cc9...",
  "ctx_hash": "da03ee8c6429da6d",
  "seed": 42,
  "scenario_hash": "2e13a520d3417572",
  "gdp_base": 1.896,
  "cpi_base": 2.095,
  "rate_base": 2.25,
  "d_gdp": -5.696,
  "d_inflation": -0.695,
  "d_interest_rate": -1.0,
  "gap_d_gdp": 0.004,
  "gap_d_inflation": 0.005,
  "gap_d_interest_rate": 0.0,
  "plausibility_ok": 1,
  "plausibility_score": 5.0,
  "regime_label_text": "stress",
  "regime_score_text": 0.6777,
  "regime_p_normal": 0.0318,
  "regime_p_stress": 0.5810,
  "regime_p_crisis": 0.3872
}
\end{lstlisting}

This record illustrates the narrative--quantitative hybrid format enforced by
the template, and shows the provenance fields---prompt hash, context hash,
template version and scenario hash---that link every reported figure back to an
immutable artefact.

\section{Prompt Templates}
\label{app:prompts}

Because prompt variant is the dominant term in every variance decomposition we
report, the full template set is a necessary part of the evidence rather than
supplementary material. All 30 templates are stored verbatim in the run artefacts
as \texttt{prompt\_templates\_tiered.json}, hashed in every run manifest under
\texttt{prompt\_template\_version}~$=$~\texttt{1d7e6dee3dc7}. The unanchored set
used by P4, \texttt{prompt\_templates\_unanchored.json}
(\texttt{02005231eafd}), contains 30 templates without a growth-shortfall band; the model still
labels a tier. Each template comprises a
narrative frame, an intended severity tier conveyed through its label and band, an indicative magnitude band, the
baseline-anchoring instructions of Section~\ref{sec:macro-inputs}, and an
identical JSON output contract. The tier assignment is: eight \emph{contained}
templates (supply, trade, energy pass-through, trade disruption, climate
transition, technology disruption, labour market, commodity price crash),
seventeen \emph{severe}, and five \emph{systemic} (credit, contagion, cyber,
emerging-market contagion, global liquidity crunch).

\section{Reproducibility Notes}
\label{app:reproducibility}

The pipeline is \emph{snapshot-replayable} rather than bit-level deterministic,
and we state the distinction precisely because it is frequently blurred.

\subsection{Frozen Retrieval and Market Snapshots}

The following artefacts are hashed with SHA-256 and recorded in a per-run
manifest: the IMF WEO workbook; serialised country profiles; the headline CSV
snapshots with their retrieval windows and query strings; MiniLM embeddings and
the persisted FAISS index; cached ETF prices and the derived PCA factors,
loadings and calm/stress covariance matrices; all prompts and parsed responses;
and the resulting risk tables and figures. Every reported number therefore
traces to a fixed set of inputs.

\subsection{Two Failure Modes That Silently Break Replay}

We document two issues encountered during development because both are easy to
miss and both would have confounded the cross-model comparison.

\paragraph{Embedding device changes the prompts.} Executing MiniLM on GPU rather
than CPU perturbs the embedding vectors at approximately $10^{-7}$, since GPU
and CPU BLAS reduce floating point in different orders. That perturbation is
harmless in isolation but feeds the $k$-means clustering that selects headline
exemplars, which is chaotic under small perturbations. In a controlled test, a
$10^{-7}$ perturbation changed four of twenty selected exemplars; a changed
exemplar rewrites the news block of a country profile, which rewrites the
prompt. Embeddings are therefore pinned to CPU execution and quantised before
any clustering or retrieval decision, and the selected exemplars are themselves
snapshotted so that clustering never re-runs on the same input.

\paragraph{News APIs are not idempotent.} Querying a news provider for a fixed
window does not return a fixed result: the endpoint returns the most recent
articles \emph{within} that window, and the set grows as more articles are
published and indexed. Two runs launched hours apart therefore receive different
headlines for a nominally identical window. Headlines are consequently fetched
once, written to disk with a recorded hash and creating run tag, and reused
verbatim by every subsequent run; the pipeline warns if snapshots originate from
more than one run tag.

\subsection{Scope of the Determinism Claim}

Given the frozen artefacts and a recorded global seed for the Monte Carlo
engine, all portfolio risk metrics regenerate up to floating-point and Monte
Carlo noise. Generation is not deterministic: the GPT-5 family does not accept
temperature below 1, and provider-side batching and kernel selection introduce
variation no client-side seed controls. As Section~\ref{sec:results-reproducibility}
documents, this yields substantial scenario-level variation alongside aggregate
stability to within $0.01\times$. We therefore claim replayability of reported
statistics, not of individual scenarios.

\subsection{Run Metadata}

Each scenario carries a minimal audit record: run identifier, country, horizon,
WEO and headline hashes, prices and PCA hashes, calm and stress-period covariance
hashes, FAISS index and MiniLM hashes, retrieval seed, prompt hash, prompt
template version, decode parameters as transmitted, scenario hash, plausibility
flags, regime label and score, and the severity index $\lambda_s$. The prompt
template version in particular allows any downstream comparison to detect that
two runs saw different prompts, which would otherwise present as a model effect.

\section{Bootstrap Confidence Intervals}
\label{app:bootstrap}

Table~\ref{tab:bootstrap-ci} reports mean linear-channel VaR and CVaR multiples
with 95\% nonparametric bootstrap confidence intervals by model, retrieval and
news configuration, based on 2{,}000 resamples of scenario-level multiples.
Intervals overlap completely across retrieval and news settings for both GPT
runs, the interval counterpart of their null ANOVA effects in
Table~\ref{tab:anova-decomp}. For Llama the news-on cells sit lower, though the
intervals still overlap, consistent with its small but significant news effect.

\begin{table*}[t]
  \centering
  \small
  \caption{Bootstrap 95\% confidence intervals for linear-channel VaR and CVaR
  multiples by model, RAG and news configuration, Portfolio~A.}
  \label{tab:bootstrap-ci}
  \begin{adjustbox}{max width=\textwidth}\begin{tabular}{lllrrrrrr}
    \toprule
    & & & \multicolumn{3}{c}{VaR multiple} & \multicolumn{3}{c}{CVaR multiple} \\
    \cmidrule(lr){4-6}\cmidrule(lr){7-9}
    Model & RAG & News & Mean & CI low & CI high & Mean & CI low & CI high \\
    \midrule
    GPT (primary) & Off & Off & 3.920 & 3.826 & 4.010 & 2.411 & 2.361 & 2.461 \\
    GPT (primary) & Off & On  & 3.952 & 3.857 & 4.050 & 2.429 & 2.377 & 2.482 \\
    GPT (primary) & On  & Off & 3.946 & 3.849 & 4.042 & 2.426 & 2.372 & 2.478 \\
    GPT (primary) & On  & On  & 3.927 & 3.830 & 4.021 & 2.415 & 2.362 & 2.467 \\
    \midrule
    GPT (repeat) & Off & Off & 3.921 & 3.827 & 4.014 & 2.413 & 2.361 & 2.465 \\
    GPT (repeat) & Off & On  & 3.954 & 3.861 & 4.051 & 2.431 & 2.380 & 2.482 \\
    GPT (repeat) & On  & Off & 3.931 & 3.835 & 4.025 & 2.417 & 2.365 & 2.469 \\
    GPT (repeat) & On  & On  & 3.957 & 3.862 & 4.053 & 2.433 & 2.382 & 2.484 \\
    \midrule
    Llama-3.1-8B & Off & Off & 4.311 & 4.171 & 4.453 & 2.626 & 2.551 & 2.703 \\
    Llama-3.1-8B & Off & On  & 4.131 & 3.979 & 4.284 & 2.529 & 2.446 & 2.613 \\
    Llama-3.1-8B & On  & Off & 4.282 & 4.145 & 4.422 & 2.613 & 2.539 & 2.686 \\
    Llama-3.1-8B & On  & On  & 4.240 & 4.082 & 4.399 & 2.587 & 2.501 & 2.675 \\
    \bottomrule
  \end{tabular}\end{adjustbox}
\end{table*}

\section{Prompt and Configuration Dispersion}
\label{app:dispersion}

\begin{table*}[t]
  \centering
  \small
  \caption{Prompt-level dispersion: for each prompt variant and configuration,
  the mean pairwise Euclidean distance between the seven countries' emitted
  (growth, inflation, rate) levels, summarised across the 30 prompt variants.
  Accepted scenarios.}
  \label{tab:prompt-dispersion}
  \begin{adjustbox}{max width=\textwidth}\begin{tabular}{lllrrrr}
    \toprule
    Model & RAG & News & Mean & Std & Min & Max \\
    \midrule
    GPT-5.6-luna & Off & Off & 1.589 & 0.488 & 0.900 & 3.205 \\
    GPT-5.6-luna & Off & On  & 1.628 & 0.493 & 0.957 & 2.943 \\
    GPT-5.6-luna & On  & Off & 1.625 & 0.519 & 0.828 & 3.046 \\
    GPT-5.6-luna & On  & On  & 1.577 & 0.470 & 0.814 & 2.788 \\
    \midrule
    GPT (repeat) & Off & Off & 1.635 & 0.539 & 0.896 & 3.051 \\
    GPT (repeat) & Off & On  & 1.595 & 0.426 & 0.910 & 2.762 \\
    GPT (repeat) & On  & Off & 1.567 & 0.485 & 0.629 & 2.855 \\
    GPT (repeat) & On  & On  & 1.553 & 0.406 & 0.842 & 2.551 \\
    \midrule
    Llama-3.1-8B & Off & Off & 3.017 & 1.150 & 1.711 & 5.982 \\
    Llama-3.1-8B & Off & On  & 3.068 & 1.268 & 1.234 & 6.630 \\
    Llama-3.1-8B & On  & Off & 2.727 & 0.874 & 0.947 & 5.180 \\
    Llama-3.1-8B & On  & On  & 3.091 & 1.356 & 1.376 & 6.830 \\
    \bottomrule
  \end{tabular}\end{adjustbox}
\end{table*}

\begin{table*}[t]
  \centering
  \small
  \caption{Intra-configuration dispersion by country (primary run, RAG and
  news both enabled), accepted scenarios, with 95\% bootstrap intervals over
  scenario pairs.}
  \label{tab:stability-config}
  \begin{adjustbox}{max width=\textwidth}\begin{tabular}{lrrrr}
    \toprule
    Country & Dispersion & CI low & CI high & $N$ \\
    \midrule
    Canada         & 3.483 & 3.344 & 3.629 & 30 \\
    France         & 3.024 & 2.912 & 3.151 & 29 \\
    Germany        & 3.304 & 3.175 & 3.432 & 30 \\
    Italy          & 3.270 & 3.123 & 3.412 & 30 \\
    Japan          & 2.905 & 2.784 & 3.021 & 30 \\
    United Kingdom & 3.420 & 3.278 & 3.566 & 30 \\
    United States  & 3.689 & 3.548 & 3.827 & 30 \\
    \bottomrule
  \end{tabular}\end{adjustbox}
\end{table*}

\end{document}